\documentclass{physeauth} 

\newcommand{\bm }{ {\overline{ m}} }
\newcommand{\del}{\delta}
\newcommand{\mb }{ {\overline{ m}} }
\newcommand{\bA}{\tilde A}
\newcommand{\bk}{\tilde k}
\newcommand{\bz}{{\bar z} }
\newcommand{\bM }{ {\overline{ M}} }
\newcommand{\Mb }{ {\overline{ M}} }
\newcommand{\bE }{ {\overline{ E}} }

\newcommand{\Ai}{{\rm Ai}}

\newcommand{\M}{\overline{\rm M}}
\newcommand{\rhoL}{{\rho_\Lambda}}

\renewcommand{\d}{{\rm d}}
\newcommand{\bi}{{\rm bi}}
\newcommand{\mub}{\bar\mu_{\rm bi}}

\newcommand{\tot}{{\rm tot}}

\newcommand{\myskip}[1]{}

\newcommand{\p}{\partial}

\newcommand{\BEQ}{\begin{eqnarray}}
\newcommand{\EEQ}{\end{eqnarray}}
\newcommand{\BEA}{\begin{eqnarray}}
\newcommand{\EEA}{\end{eqnarray}}
\newcommand{\nn}{\nonumber }
\renewcommand{\d}{{\rm d}}

\renewcommand{\ni}{\noindent}
\newcommand{\mn}{{\mu\nu}}

\newcommand{\lambdaB}{{\bar\lambda}}
\newcommand{\lambdaBB}{{\tilde\lambda}}

\begin{document}

\begin{frontmatter}

\title{ Bose-Einstein condensed supermassive black holes: a case of
renormalized quantum field theory in curved space-time}
\author[UvA]{Theo M. Nieuwenhuizen\thanksref{thank1}}

\address[UvA]{Institute for Theoretical Physics, Valckenierstraat 65,
 1018 XE Amsterdam, The Netherlands}
\thanks[thank1]{e-mail: t.m.nieuwenhuizen@uva.nl}

\author[TAc]{V. \v Spi\v cka \thanksref{thank2}}
\address[TAc]{Institute of Physics, Academy of Sciences
of the Czech Republic, Na Slovance 2, 182 21 Praha
8, Czech Republic}
\thanks[thank2]{e-mail: spicka@fzu.cz}

\begin{abstract}
This paper investigates the question whether a realistic black
hole can be in principal similar to a star, having  a large but
finite redshift at its horizon. If matter spreads
throughout the interior of a supermassive black hole with mass
$M\sim10^9M_\odot$, it has an average density comparable
to air and it may arise from a Bose-Einstein condensate of densely
packed H-atoms. Within the Relativistic Theory of Gravitation with
a positive cosmological constant, a bosonic quantum field
describing H atoms is coupled to the curvature scalar with
dimensionless coupling $\xi$. In the Bose-Einstein condensed
groundstate an exact, self-consistent solution for the metric
occurs for a certain large value of $\xi$, quadratic in the
black hole mass. It is put forward that $\xi$ is set by proper choice 
of the background metric as a first step of a renormalization approach,
while otherwise the non-linearities are small. The black hole has a hair,
the binding energy. Fluctuations about the ground state are considered.
\end{abstract}

\begin{keyword}
supermassive black hole\sep quantum field theory\sep
Bose Einstein condensation\sep renormalization 
\PACS 04.70.Bw\sep 04.20.Cv \sep 04.20.Jb
\end{keyword}

\end{frontmatter}

\section{Introduction}

The standard knowledge that black holes (BHs) have no hair (i. e., they are determined
by mass, charge and spin only) has recently been tested versus observational data 
from quasars.  Schild et al. carefully examine why 
standard models of quasars with the various structural elements fail to be 
compatible with the observations of a luminous ring at the innermost edge of the 
accretion disc of the quasar Q0957+561 A,B. 
The observations imply that this quasar contains an observable magnetic moment,
 a ``hair'', which represents strong evidence that it does not have an event horizon 
~\cite{Schildetal06} . On a different track, one of us presented a class of exactly solvable
BHs of with the Schwarzschild on the outside having with one hair, 
namely the binding energy ~\cite{NEPLBH08}.

A great deal of observational activity is presently underway to measure
with 10 micro-arcsec resolution the Sgr A black hole object at the center
of our Galaxy, because it is widely claimed that detecting the shadow of
the black hole would prove the existence of an event horizon. However the
black hole with hair objects either indicated by the Schild et al (2006) 
observations ~\cite{Schildetal06} or by the Nieuwenhuizen (2008) theory~\cite{NEPLBH08},
would also be a compact object with a strong gravity shadow
that is virtually indistinguishable from a black hole event horizon. 
Indeed, the outer metrics are close to the Schwarzschild metric.
The two classes of objects are in fact best distinguished from one another by the
existence of a light cylinder effect originating the dusty torus, and the
clear demonstration of such light cylinder effects in ordinary quasars
by Schild et al (2009) favors the magnetic over the standard black hole
~\cite{Schildetal09}.

Quasars are probably supermassive black holes which occur at the center of most  galaxies. 
Even though the one of our Galaxy weighs only about 4
million solar masses, the typical weight is a billion solars, with
the present champion at 17 billion. Since the Schwarzschild radius of a black
hole scales with its mass, $r_S=2GM/c^2$, its density decays as
$1/M^2$, implying that bigger black holes have a lower average
density. This motivates to consider a supermassive black hole, say
with $M=10^9M_\odot$. Then the average mass density is on the order of grams
per liter, comparable to air. The fact that they are enormously heavy
only because of their size,  suggests that for these
objects the physics may be not too difficult or unfamiliar, and it
invites to study their internal structure as a standard  problem 
in a standard theory.

These estimates of the average density confront theoretical
descriptions. On the basis of the Schwarzschild, Kerr and
Kerr-Newman metrics, black holes (BHs) are described as singular
objects with all matter localized in the center or, if rotating,
on an infinitely thin ring. Recent approaches challenge this
assumption and consider matter just spread throughout the
interior~\cite{Laughlin,Dymnikova,MazurMottola}.

Let us point at the following simple connection. 
Supermassive BHs occur in the center of many galaxies and weigh about $M_{\rm
BH}=0.0012\,M_{\rm bulge}$ \cite{BHmassBulgeMass}. Let us assume
that they consist of hydrogen atoms and that mass and particle
number are related as $M\equiv \nu Nm_{\rm N}$ with $m_N$ the nucleon mass and 
some $\nu\le 1$. Neglecting rotation, we may compare the H number density within the 
Schwarzschild radius $R_S=2GM/c^2$, i.e.
$n_H=3N/4\pi r_S^3$ with the one of densely packed,
non-overlapping H-atoms, that is, with the Bohr density 
$n_{\rm B}\equiv 3/4\pi a_0^3$, with $a_0=0.529\,$\AA the Bohr radius.
This yields a mass $c^3(a_0^3/8G^3m_N)^{1/2}=8.26\cdot 10^7 M_\odot$ 
(we take $\nu=1$ here), which indeed lies in the range of observed
supermassive black holes. 
The corresponding mass density is $2.7$ kg/liter, but scaling as $1/M^2$ 
for larger black holes, it  is as low as  0.6 g/liter for the $17\cdot 10^9M_\odot$ 
black hole, already below the one of  air. We shall therefore take as characteristic
mass scale $10^9M_\odot=1.98\cdot 10^{39}$ kg and write

\BEQ M\equiv M_9\cdot10^9M_\odot. \EEQ


The assumption of spread-out matter poses the question how matter
can have a pressure that allows such a state. It was proposed
originally by Sacharov that the vacuum equation of state $p=-\rho$
could describe matter at superhigh densities~\cite{Sacharov}.
Laughlin and coworkers assume that matter near the horizon could
be in its Bose-Einstein condensed (BEC) phase, modeled by the
vacuum equation of state ~\cite{Laughlin}. Dymnikova considers BHs
obeying it in the interior, which, however, have one or two
horizons ~\cite{Dymnikova}. Mazur and Mottola take the BEC idea
over to the interior, and investigate a ``gravo-star'', of which
the interior obeys the vacuum equation of state, and which is
surrounded by a thin shell of normal matter having the stiff
equation of state $p=+\rho$. This solution is regular
everywhere~\cite{MazurMottola}.  A related subject is the
description of boson stars ~\cite{bosonstar1,bosonstar2}.

One of us considered a
modification of general relativity (GR), the Relativistic Theory
of Gravitation (RTG) that differs from GR only at very high
redshift, in particular near the horizon of black holes~\cite{NEPLRTG07}. 
This modification appears to describe the BH interior like a star, 
be it with strong but finite redshifts.

We shall study a supermassive BH that exist as a self-gravitating
hydrogen cloud, in a Bose-Einstein condensed phase.  Hereto we
employ the Relativistic Theory of Gravitation (RTG), which reproduces 
all weak gravitational effects in the solar system ~\cite{LogunovBook,LogInflaton} 
as well as the $\Lambda$CDM cosmology ~\cite{NEPLRTG07}.
We shall extend the recent approach to this line of research
started by one of us~\cite{NEPLBH08,NAIPBH08,NFNL08},
where an exact interior solution to the supermassive black hole problem was found,
in a theory which is a modification of  General Relativity. 
This Schwarzschild-type black hole has a hair, namely the binding energy 
of the matter out of which it is composed.  It has a very large but finite 
redshift at the horizon, so there is no sharp horizon, 
no Hawking radiation and no role for Bekenstein-Hawking entropy.
From a principle point of view, this approach, when extended to the rotating
situation with an accretion disk, may describe the above discussed quasar observations.

Section 2 discusses aspects of RTG and its generalization of the 
Schwarzschild metric, together with its deformation near the horizon.
Section 3 introduces the quantum field theory for the H-atoms
and their Bose-Einstein condensation.
Section 4 presents an exact solution of the interior metric.
Section 5 deals with general solution near the horizon.
Section 6 discusses excitation about the groundstate and 
the paper closes with a discussion in section 7.

\section{Relativistic Theory of Gravitation}

We consider a static metric $\d{\rm s}^2=g_\mn\d x^\mu\d x^\nu$,
where $x^\mu=(ct,r,\theta,\phi)$, with $\mu=0,1,2,3$.
Here and in the sequel, summation over repeated indices is implied. 
In case of spherical symmetry it has the form

\BEQ \label{sphersymmetric}
\d {\rm s}^2=U(r)c^2\d t^2-V(r)\d r^2-W^2(r)\d\Omega^2\EEQ

\ni with $\d\Omega^2=\d\theta^2+\sin^2\!\theta\d\phi^2$.
The metric tensor can be read off, $g_\mn={\rm diag}(U,-V,-W^2,-W^2\sin^2\theta)$,
while $g^\mn=(g^{-1})_\mn$.
The gravitational energy density arises from the Landau-Lifshitz pseudo-tensor
\cite{LandauLifshitz}, generalized to become a tensor in Minkowski space-time
~\cite{BabakGrishchuk,NEPLRTG07}.
For the metric (\ref{sphersymmetric}) it takes the form
\BEQ \label{t00r=}
t^{00}&=&\frac{c^4W^2}{8\pi Gr^6}
\left(-\frac{r^2V'WW'}{V}+r^3V' -5r^2W'{}^2\right.\nn\\
&+& \left.\frac{2r^3VW'}{W}+8rWW'-2r^2V-3W^2\right). \EEQ

Let us start with the General Theory of Relativity (GTR, GR). The
Schwarzschild metric reads in the harmonic gauge

\BEQ
\label{Schwarzharm}
U_S=\frac{1}{V_S}=1-\frac{2\bM}{W_S}=\frac{r-\bM}{r+\bM},\quad W_S=r+\bM,
\EEQ 

\ni where $\bM$ is the gravitational length

\BEQ \bM=\frac{GM}{c^2}.\EEQ

\ni The metric is singular at the horizon
$W_S=2\bM$, $r=\bM$ and it involves the gravitational energy
density

\BEQ \label{ThetaSS} t^{00}=\frac{c^4\bM}{4\pi G r^2}\frac{\d}{\d
r}\frac{(r+\bM)^3(2r+\bM)}{2r^3(r-\bM)}. \EEQ

\ni Its quadratic divergence
at $\bM$ presents an often overlooked, physical peculiarity, that
induces a negative infinite contribution to the total energy. In other gauges
the singularity can be moved completely to the origin. All by all,
the situation is puzzling. For
this reason, we shall switch to RTG with matter not located at the
singularity $r=0$, but just spread out within the horizon.

We focus on RTG, which describes gravitation as a field in Minkowski space
 ~\cite{LogunovBook,LogInflaton} and possesses the same gravitational energy
momentum tensor and thus also the gravitational energy density Eq.
(\ref{t00r=})~\cite{NEPLBH08}. It extends the Hilbert-Einstein action
with the cosmological term and a bimetric coupling between the
Minkowski ($\gamma$) and Riemann ($g$) metrics,

\BEQ \label{Lagtot=}
L=-\frac{c^4R}{16\pi G}-\rho_\Lambda+\half\rho_\bi \gamma_\mn g^\mn+L_{\rm mat},
\EEQ

\ni (For $\rho_\bi=0$ it is just a field theoretic description of GTR.)
The dimensionless action is $S/\hbar=(1/\hbar c)\int\d^4x\sqrt{-g}L$,
where $x_0=ct$.  One has the Einstein equations

\BEQ \label{EinEqs}
&& G^\mn\equiv R^\mn-\half g^\mn R=\frac{8\pi G}{c^4}T_{\rm tot}^\mn, \\&&
T_{\rm tot}^\mn=T^\mn+T_\Lambda^\mn+T_\bi^\mn,\EEQ

\ni where $T^\mn$ comes from the matter, $T_\Lambda^\mn$
from the cosmological term and $T_\bi^\mn$ from the bimetric term.
They have the dimension of energy/volume
and elements $(\rho,-p_i)$ of the form

\BEQ \label{rhotot=}
\rho_{\rm tot}&=&\rho+\rhoL
+\frac{\rho_\bi}{2U}-\frac{\rho_\bi}{2V}-\frac{\rho_\bi r^2}{W^2},\nn\\
p_{i}^{\rm tot}&=&
p_{i}-\rhoL+\frac{\rho_\bi}{2U}-\frac{\rho_\bi}{2V}+\frac{\rho_\bi
r^2}{W^2}. \EEQ

\ni with $i=r,\theta,\phi$.  The mass density is $\rho/c^2$.
 It is custumary to choose the value
$\rho_\bi=\rho_\Lambda$ in order to allow a Minkowski metric $U=V=1$, $W=r$ in
the absence of matter, because then $\rho_{\rm tot}=p^{\rm tot}_i=0$.
We shall fix $\rhoL=\rho_\bi$ to the observed positive
cosmological constant~\cite{NEPLBH08}. However, historically the
opposite choice $\rho_\Lambda<0$ was considered and the
cosmological data were described by an additional inflaton
field ~\cite{LogInflaton}, so the sign of $\rho_\bi$ is still disputed. 
We show that the possibility to solve a realistic black hole
settles that indeed $\Lambda_\bi>0$ is the physically interesting case.
The new point of RTG is that $g_{00}=U$ can be very small but still positive.
Despite the smallness of $\rho_\bi$, the $\rho_\bi/U$ term becomes relevant near the
horizon~\cite{LogunovBook,NEPLBH08} and regularizes the singularities of
the Schwarzschild metric. So we continue with $\rho_{\rm tot}\approx \rho+\rho_\bi/2U$,
$p_i^{\rm tot}\approx p_i+\rho_\bi/2U$.

The $\rho_\bi$ term in (\ref{Lagtot=}) violates general coordinate invariance
and the consistency requirement $T^\mn_{\bi\,;\nu}=0$ imposes 
the harmonic gauge condition

\BEQ \label{harm}
\frac{U'}{U}-\frac{V'}{V}+4\frac{W'}{W}=\frac{4rV}{W^2}.
\EEQ

\ni The residual gauge group actually would lead to an infinite
gravitational energy, so its physical subgroup is empty, making
the solution unique~\cite{NRTGlong}.

In post-Newtonian approximations and in applications to, e. g.,  stars, 
the $\rho_\Lambda$ and $\rho_\bi$ terms are negligible, which will bring
back the general coordinate invariance of GTR. Gravitational radiation
in e. g. X-ray binaries is the same as in GTR. But near the horizon of a 
black hole the large redshift will make the $\rho_\bi/U$ term sizeable,
and deeply change the theory.

The $^0_0$, $^1_1$ and $^2_2$ components of the Einstein equations read
respectively 
 \BEQ\label{Ein00}
&&\frac{1}{W^2}-\frac{W'{}^2}{VW^2} -\frac{2W''}{VW}
+\frac{V'W'}{V^2W}=\frac{8\pi G}{c^4} \rho_{\rm tot},
\nn \\
\label{Ein11} && \frac{1}{W^2}-\frac{W'{}^2}{VW^2}
-\frac{U'W'}{UVW} =-\frac{8\pi G}{c^4}p_{r}^{\rm tot}, \\
\label{Ein22} &&
-\frac{U''}{2UV}-\frac{W''}{VW}+\frac{U'{}^2}{4U^2V}+\frac{U'V'}{4UV^2}
+\frac{V'W'}{2V^2W} \nn\\&&-\frac{U'W'}{2UVW}=
-\frac{8\pi G}{c^4}p_{\perp}^{\rm tot} \nn, \EEQ 

\ni As usual, the last equation is automatically satisfied by energy 
conservation $T^\mn_{;\nu}=0$. The Ricci scalar becomes

\BEQ R&=&\frac{8\pi G}{c^4}(-\rho_{\rm tot}+p_r^{\rm tot}+2p_\perp^{\rm tot})\equiv R_\bi+R_m
\\ R_\bi&=&\frac{8\pi G}{c^4}\frac{\rho_\bi}{U},\quad
R_m=\frac{8\pi G}{c^4}(-\rho+p_r+2p_\perp).\EEQ

We define the inverse length $\mu_\bi$ by

\BEQ \rho_\bi=\frac{c^4\mu_\bi^2}{16\pi G},\qquad \mu_\bi=\sqrt{2\Lambda}=
2.38\cdot10^{-23}\frac{c^2}{GM_\odot},\EEQ

\ni with $\Lambda$ the cosmological constant. We shall often encounter the combination

\BEQ \mub=\frac{\mu_\bi GM}{c^2}= 2.38\cdot10^{-23}\frac{M}{M_\odot}
=2.38\cdot 10^{-14}{M_9}.\EEQ


\subsection{Generalization  of the Schwarzschild solution}

The Schwarzschild metric solves the above equations in the limit
$\rho\to0$, $\rho_\Lambda=\rho_\bi\to0$. A more general solution
within RTG is found as follows. Assume that $U$, $V$ and $W$ are
still related in the Schwarzschild manner $U=1-2\bM/W$, $V=W'{}^2/U$.
This solves the two Einstein equations (\ref{Ein00}) without matter,
$\rho_\tot=p_\tot=0$. Inserting this in the harmonic constraint we
have

\BEQ -\frac{W''}{W'{}^3}W(W-2\bM)+\frac{2(W-\bM)}{W'}=2r \EEQ
Going to the inverse function $r(W)$ this becomes linear,
\BEQ W(W-2\bM)r''+2(W-\bM)r'=2r \EEQ
One solution is $r=A(W-\bM)$.
The method of variation of constants brings


\BEQ \frac{A''}{A'}=-\frac{1}{W}-\frac{2}{W-\bM}-\frac{1}{W-2\bM} \EEQ
So with integration constant $C$ one has

\BEQ A'=\frac{-C\bM^3}{W(W-\bM)^2(W-2\bM)}.
\EEQ

\ni It brings $A$ and from this, choosing $A(\infty)=1$,
~\footnote{An exact solution was found before by A. V. Genk and A. A. Tron.}

\BEQ\label{rass=}\label{genSschws}
r=W-\bM+C\left[\frac{W-\bM}{2}\log\frac{W}{W-2\bM}-\bM\right]. \EEQ 

\ni The horizon $U=0$, $W=2\bM$ is located at $r=\infty$ for $C>0$
and at $-\infty$ for $C<0$.
Clearly, a small $C$ is expected to avoid such peculiarities.

Near the horizon the bimetric coupling induces a deformation of the 
Schwarzschild metric, which regularizes its singularity
~\cite{LogunovBook,LogInflaton}. For small values of the dimensionless product 
$\mu_\bi \bM$, a scaling form was presented by one of us \cite{NEPLBH08},

\BEA\label{scaling} r&=&M
\frac{1+\eta(e^\zeta+\zeta+\log\eta+2)}{1-\eta(e^\zeta+\zeta+\log\eta+2)},
\qquad
 U = \eta e^\zeta, \\
V&=&\frac{e^ \zeta}{\eta(1+e^\zeta)^2},\qquad W=\frac{2M}{1- \eta
e^\zeta-\mub^2(\zeta+w_0)}.\nn \EEA

\ni Here $\zeta$ is the running variable and $\eta\sim\mu_\bi \bM$ a
small scale. For $\mu_\bi\bM\ll\eta e^\zeta\ll 1$ it coincides with the generalized
Schwarzschild solution (\ref{genSschws}), with $C=-4\eta$ indeed being small. 
Thus the singularities of (\ref{genSschws}), $U$ and $V$  are 
smoothly deformed in (\ref{scaling}), due to the bimetric term $\rho_\bi/U$.

\section{ Quantum field theory in curved space}

Let our H-atoms be described by a scalar, bosonic creation field operator
~\footnote{In field theory in flat space one usually adds a factor 
$1/\sqrt{2E_i}$ in the terms of (\ref{fieldsum}); in curved theory space 
these factors are moved to the inner product, see Eq. (\ref{innerprod}).~\cite{BirrellDavies}} 

\BEQ\label{fieldsum} \hat\psi({\bf r},t)=\sum_i \hat a_i\psi_i({\bf r})e^{-iE_it/\hbar},  \EEQ

\ni where $i=\{n,\ell,m\}$, $[\hat a_i, \hat a_j^\dagger]=\delta_{ij}$
and eigenfunctions factor as $\psi_i({\bf r})=\phi_{n\ell}(r)$$Y_{\ell
m}(\theta,\phi)$. The total field ~\cite{BirrellDavies}

\BEQ \hat\psi_{\rm t}=\hat\psi+\hat\psi^\dagger
\EEQ

\ni  is Hermitean. We assume that
 the two-particle interaction can be replaced
by a $\delta$-potential. This leads to a quartic Lagrangian density

\BEQ \label{Lagmfull} L_{\rm mat}^{\rm full}=\half 
\p^\mu\hat
\psi_{\rm t}\p_\mu\hat\psi_{\rm t} - (\frac{m^2c^2}{2\hbar^2}
+\frac{\xi}{2} R)\hat\psi_{\rm t}^2
-\frac{\lambda}{24\hbar c}\hat\psi_{\rm t}^4. \EEQ

\ni where $\p_\mu=\p/\p x^\mu$, $\p^\mu=g^\mn\p_\nu$ and $\lambda$
is dimensionless. With field dimension $[\psi]=[\sqrt{\hbar c}/ \ell_P]$,
where 

\BEQ \ell_P=\sqrt{\frac{\hbar G}{c^3}},\quad m_P=\sqrt{\frac{\hbar c}{G}},
\EEQ

\ni are the Planck lenght and the Planck mass, respectively, the Langrangian
density has dimension $[ L_{\rm mat}^{\rm full}]=[\hbar c/\ell_P^4]$, 
i. e., energy  density, as it should.

For a field in curved space the renormalization group generates
the coupling to the Ricci curvature scalar $R$ entering eq.
(\ref{Lagmfull})~\cite{BirrellDavies}. Its strength $\xi$ has to be
obtained from renormalization arguments. Popular values are
$\xi=0$ and $\xi=\frac{1}{6}$. In our earlier work
~\cite{NEPLBH08} we have treated it as a phenomenological
parameter, here we shall argue that its large, mass dependent value is
self-generated.

The conjugate momentum field is

\BEQ \hat\pi_t^\mu=\frac{\partial L_{\rm mat}^{\rm full}}
{\partial (\p_\mu\hat\psi_{\rm t})}=\,\p^\mu\hat\psi_{\rm t}.
\EEQ

\ni The equal-time commutation relation is for the field is

\BEQ n_\mu[\hat\psi_{\rm t}({\bf r},t),\hat\pi_{\rm t}^\mu({\bf r'},t)]=
i\hbar c\frac{\delta^{(3)}({\bf r}-{\bf r'})}{\sqrt{-g_3}},
\EEQ

\ni where $n_\mu=\delta^0_\mu\sqrt{U}$ is a time-like unit vector and
$g_3=-VW^4\sin^2\theta$ the determinant of the spatial part of the
metric. The Hamiltonian density is

\BEQ &&\hat H_{\rm t}=\hat\pi_{\rm t}^0\p_0\hat\psi_{\rm t}-L_{\rm mat}^{\rm full}=\\
&&
\frac{\p^0\hat\psi_{\rm t}\p_0\hat\psi_{\rm t} +\p^i\hat\psi_{\rm t}\p_i\hat\psi_{\rm t} }{2}
+(\frac{m^2c^2}{2\hbar^2}+\frac{\xi}{2} R)\hat\psi_{\rm t}^2
+\frac{\lambda}{24\hbar c}\hat\psi_{\rm t}^4. \nn\EEQ

\ni where $i=1,2,3$ sums the spatial components.

Bose-Einstein condensation can be described in the rotating wave approximation,
yielding, after normal ordering, the Lagrangian density
~\cite{PitaevskiiStringari}

\BEQ \label{Lagm=} L_{\rm mat}=g^\mn \p_\mu\hat
\psi^\dagger\p_\mu\hat\psi -(\mb^2+\xi R)\hat\psi^\dagger\hat\psi
-\frac{\lambda\hat\psi^\dagger{}^2\hat\psi^2}{4\hbar c}, \EEQ

\ni where we define the inverse Compton length

\BEQ \mb=\frac{mc}{\hbar}. \EEQ

\ni The equal-time commutator reduces to

\BEQ \label{comrel=}
&&n^0[\hat\psi({\bf r},t),\p_0\hat\psi^\dagger({\bf r'},t)]
+n^0[\hat\psi^\dagger({\bf r},t),\p_0\hat\psi({\bf r'},t)]\nn\\
&&= i\hbar c\frac{\delta^{(3)}({\bf r}-{\bf r'})}{\sqrt{-g_3}},
\EEQ
\ni which reads when written out in eigenfunctions 

\BEQ \label{orthonorm}
n^0\sum_j\frac{2iE_j}{\hbar c}\psi_j({\bf r})\psi_j^\ast({\bf r'})= 
i\hbar c\frac{\delta^{(3)}({\bf r}-{\bf r'})}{\sqrt{-g_3}}.
\EEQ

\ni Multiplying this with $\sqrt{-g_3({\bf r})}\psi_i^\ast({\bf r})\d^3r$ leads to the orthonormality  
$ (2E_i/\hbar^2c^2)\int\d^3r\,n^0\sqrt{-g_3}\psi_i^\ast\psi_j=\delta_{ij}$.
The Hamiltonian density becomes 
\BEQ 
\hat H&=&\p^0\hat\psi^\dagger\p_0\hat\psi +\p^i\hat\psi^\dagger\p_i\hat\psi 
+(\bm^2+\xi R)\hat\psi^\dagger\hat\psi\nn\\
&+&\frac{\lambda}{4\hbar c}\hat\psi^\dagger{}^2\hat\psi^2. \EEQ


The non-relativistic, flat space Gross-Pitaevskii equation reads~\cite{PitaevskiiStringari}

\BEQ \label{GPeq0}
i\hbar\p_t\Psi_0= -\frac{\hbar^2}{2m}\nabla^2\Psi_0+g|\Psi_0^2|\Psi_0,\EEQ

\ni with $g=4\pi\hbar^2 a_s/m$ modeling the two particle interaction by the
scattering length $a_s$. For hydrogen in flat space one has~\cite{Stoof}

\BEQ a_s=0.32\,a_0\quad\textrm {singlet state},\qquad
 a_s=1.34a_0 \quad \textrm {triplet}. \nn \EEQ

\ni We shall continue with the singlet value.
Our first task is to connect $\lambda$ to $g$. The relativistic form of (\ref{GPeq0}) is

\BEQ \label{GPflatrel}
\eta^\mn \p_\mu\p_\nu\psi_0 +\mb^2\psi_0
+\frac{\lambda N_0}{2\hbar c}|\psi_0|^2\psi_0=0,\EEQ

\ni where $\eta^\mn={\rm diag}(1,-1,-1,-1)$ is the Minkowski metric and
$\psi_0=\hbar c\Psi/\sqrt{2N_0mc^2}$, while the coupling,

\BEQ\label{lambda=}
 \lambda=\frac{8m^2cg}{\hbar^3}=32\pi\frac{a_smc}{\hbar }=32\pi a_s\bm
= 8.10\cdot10^6,
\\ 
\EEQ

\ni is dimensionless. Eq. (\ref{GPflatrel}) derives from the groundstate
of a field (\ref{fieldsum}) with $N_0$ particles in the groundstate 
of a field theory (\ref{Lagm=}) with metric $g^\mn=\eta^\mn$ and $\xi=0$. 
Going to curved space is achieved by taking a general $g^\mn$, which yields 
the bare Lagrangian density, still having $\xi=0$ and $\lambda$ from (\ref{lambda=}).
Renormalization arguments will add the $\xi R$ term, and we shall choose $\xi$ such that 
the Bose-Einstein condensed problem has a self-consistent solution.
Thus, it is assumed that $\xi$ is set self-consistently, and it may consequently
depend on BH parameters such as the mass.

\myskip{
Gross-Pitaevskii equation in flat space derives from a Lagrangian density

\BEQ \label{Lagm333=} \eta^\mn \p_\mu\hat
\psi^\dagger\p_\nu\hat\psi -\mb^2\hat\psi^\dagger\hat\psi
-\frac{\lambda}{4\hbar c}\hat\psi^\dagger{}^2\hat\psi^2, \EEQ

where $\hat\psi$ has the vacuum expectation value $\hat\psi=\hbar c\Psi/\sqrt{mc^2}$.

curved  space one may attempt to go to a local inertial frame and transforming 
back to an arbitrary frame. It therefore reads

\BEQ \label{GPeq}
\left(\frac{1}{\sqrt{-g}}\p_\mu \sqrt{-g}g^\mn\p_\nu+\bm^2+\xi R+
\frac{\lambda|\Psi_0^2|}{4\bE_0}\right)\Psi_0 =0.
\EEQ

\ni with inverse length $\bE_0$,

\BEQ 
\bE_0\equiv\frac{E_0}{\hbar c}.
\EEQ

\ni The dimensionless coupling $\lambda$ reads in general

\BEQ \lambda=\frac{8E_0^2g}{\hbar^3c^3}.
\EEQ

\ni In our previous approach we took here $E_0\to mc^2$~\cite{NEPLBH08},
but we shall see that it leads to absurdly large corrections to the groundstate.
Therefore we shall insert below the actual, much smaller value for $E_0$.
}


\subsection{Self-consistent field theory}

The material energy-momentum tensor
$(T_m)^\mu_\nu\equiv(\rho_m,-p^m_r,-p^m_\theta,-p^m_\phi)$ is
derived from the quantum field theory. We do not include
the effect of the $\xi R$-term in it. The energy density reads

\BEQ \label{rhopgen=}
\rho_m&=&\frac{\langle\p_t\hat\psi^\dagger\p_t\hat\psi\rangle}{c^2U}
+\frac{\langle\p_r\hat\psi^\dagger\p_r\hat\psi\rangle}{V}
+\frac{\langle\p_\theta\hat\psi^\dagger\p_\theta\hat\psi\rangle}{W^2}\nn\\
&+&
\frac{\langle\p_\phi\hat\psi^\dagger\p_\phi\hat\psi\rangle}{W^2\sin^2\theta}
+\bm^2\langle\hat\psi^\dagger\hat\psi\rangle
+\frac{\lambda}{4\hbar c}\langle\hat\psi^\dagger{}^2\hat\psi^2\rangle.
\EEQ

\ni The pressures $(p^m_{r},p^m_{\theta},p^m_{\phi})$ have this shape
with signature $(++----)$, $(+-+---)$, and $(+--+--)$, respectively.
Spherical symmetry will imply that $p^m_\theta=p^m_\phi\equiv p^m_\perp$.
For a uniform groundstate $p_m$ is isotropic,

\BEQ \label{rhopground=}
(\rho_m,p_m)=\frac{E_0}{2U}|\Psi_0^2|
\pm\left(\frac{m^2}{2E_0}|\Psi_0^2|+\frac{\lambda|\Psi_0^4|}{16E_0^2}
\right).\EEQ

\ni They consist of a vacuum part $p=-\rho=$ const.
and a stiff part $p=+\rho\sim1/U$, the types studied in
~\cite{MazurMottola} and in \cite{NFNL08}.
In the non-relativistic ($E_0=mc^2$) and flat space ($U=1$) limit,
they reduce for $\lambda=0$ to the expected results 
$\rho_m=mc^2|\Psi_0^2|$ and $p_m=0$.

The energy momentum tensor in the Einstein equations has
two further terms. Because of the $\xi R$-term in (\ref{Lagm=}),
the Einstein equations embody a {\it direct backreaction} ($B$) of matter on
curvature, $G^\mn=8\pi Gc^{-4}(T^\mn_m+T^\mn_\Lambda+T^\mn_\bi+T_B^\mn)
-B G^\mn$, with  

\BEQ \label{Bdef} B&=&\frac{16\pi G}{c^4}\xi\langle\hat\psi^\dagger\hat\psi\rangle\equiv 
B_0+B_e,\\
B_0&=&\frac{16\pi G}{c^4}\xi N|\psi_0|^2,\quad 
B_e=\frac{16\pi G}{c^4}\xi\langle\hat\del\psi^\dagger\del\hat\psi\rangle.\nn
\EEQ

\ni 
If $B$ depends on $r$, the derivation of the Einstein
equations brings from the term $\xi R\langle\hat\psi^\dagger\hat\psi\rangle$ 
also derivatives of $B$, which induces another new term $T^\mn_B$.
Let us in general define the dimensionless density and pressures

\BEQ \bar\rho=\frac{8\pi G}{c^4}\bM^2\rho,\quad
\bar p_i=\frac{8\pi G}{c^4}\bM^2p_i.
\EEQ

\ni The elements of $(T_B)^\mu_\nu\equiv {\rm
diag}(\rho_B,-p^B_r,-p^B_\perp,-p^B_\perp)$ can then be expressed as

\BEQ \bar\rho_B&=&
\bM^2\left(\frac{B''}{V}+\frac{2rB'}{W^2}-\frac{B'U'}{2UV}\right)
,\nn\\
\bar p^B_r&=&
-\bM^2\left(\frac{B'U'}{2UV}+\frac{2B'W'}{VW}\right), \\
\bar p^B_\perp
&=&
-\bM^2\left(\frac{B''}{V}+\frac{2rB'}{W^2}-\frac{B'W'}{VW}\right).\nn
\EEQ

\ni where the explicit factors $r$ arise from inserting the
harmonic constraint (\ref{harm}). 

In terms of the function $B_0$, eq. (\ref{rhopgen=})  and the pressures read

\BEQ \rho_m&=&\frac{\bE_0^2B_0}{U}+\frac{B_0'{}^2}{4VB_0}+\frac{\bm^2}{16\pi\xi}
(B_0+\frac{\lambdaB}{2} B_0^2)+\rho_e,\\
p_r^m&=&\frac{\bE_0^2B_0}{U}+\frac{B_0'{}^2}{4VB_0}-\frac{\bm^2}{16\pi\xi}
(B_0+\frac{\lambdaB}{2} B_0^2)+p_r^e,\\
p_\perp^m&=&\frac{\bE_0^2B_0}{U}-\frac{B_0'{}^2}{4VB_0}-\frac{\bm^2}{16\pi\xi}
(B_0+\frac{\lambdaB}{2} B_0^2)+p_\perp^e,\\
\EEQ

\ni where 

\BEQ 
\bE_0\equiv\frac{E_0}{\hbar c},
\EEQ

\ni and where $\rho_e$, $p_r^e$ and $p_\perp^e$ arise from the excited states.
They are derived in (\ref{repeTe}).
Together with $B_e$, they represent all fluctuations of the problem
about the ground state at temperature $T>0$; 
in fact, even at $T=0$  these fluctuations exist and they already bring important effects, 
see ~\cite{PitaevskiiStringari} in general and the discussion below for our BH.
Furthermore, the $B_0^2\sim|\psi_0^4|$ term is of relative order

 \BEQ \label{lambdaB0=}
\lambdaB \equiv \frac{\lambda m_P^2}{32\pi m^2\xi} .
\EEQ

Eq. (\ref{EinEqs}) involves a total
energy momentum tensor 

\BEQ \label{TmnBr} T^\mn_{\rm tot}
=\frac{T_m^\mn+T^\mn_\Lambda+T_\bi^\mn+T_B^\mn}{1+B}\equiv
T^\mn+T^\mn_\Lambda+T^\mn_\bi, \nn\\ \EEQ

\ni with elements $(T_{\rm tot})^\mu_\nu\equiv{\rm diag}(\rho_{\rm tot},
-p^{\rm tot}_r,-p^{\rm tot}_\perp,-p^{\rm tot}_\perp)$.
As mentioned, of the last two terms in (\ref{TmnBr}) only the $\rho_\bi/U$ contributions
are relevant for us.
$T^\mu_\nu\equiv{\rm diag}(\rho,-p_r,-p_\perp,-p_\perp)$ has dimensionless elements

\BEQ \label{rhoB=}
 \bar\rho&=&
 \frac{\bM^2}{1+B}
 \left[ \frac{B''}{V}+\frac{2rB'}{W^2}-\frac{B'U'}{2UV}
+\frac{B_0'{}^2}{8\xi B_0V}\right.\nn\\&&\left.
+\frac{\bm^2}{2\xi}B_0 +\frac{\bm^2}{4\xi}\lambdaB B_0^2
+\frac{2\bE_0^2B_0- \mu_\bi^2\xi B}{4\xi U}+\frac{\bar \rho_e}{\bM^2}\right],  \nn\\
\label{prB=}
\bar p_r&=& \frac{\bM^2}{1+B}\left[
-\frac{B'U'}{2UV} -\frac{2B'W'}{VW} +\frac{B_0'{}^2}{8\xi B_0V}\right.  \\&&\left.
-\frac{\bm^2}{2\xi}B_0 -\frac{\bm^2}{4\xi}\lambdaB B_0^2
+\frac{2\bE_0^2B_0- \mu_\bi^2\xi B}{4\xi U}+\frac{\bar p_r^e}{\bM^2}\right],  \nn\\
\label{pperpB=}
\bar p_\perp&=&
  \frac{\bM^2}{1+B}\left
[-\frac{B''}{V} -\frac{2rB'}{W^2}+\frac{B'W'}{VW}
-\frac{B'{}^2}{8\xi BV}\right.\nn\\&&\left.
-\frac{\bm^2}{2\xi}B_0 -\frac{\bm^2}{4\xi}\lambdaB B_0^2
+\frac{2\bE_0^2B_0- \mu_\bi^2\xi B}{4\xi U}+\frac{\bar p_\perp^e}{\bM^2}\right]. \nn
\EEQ


The Ricci scalar follows from
$R={\mu_\bi^2}/{2U}+R_m$, where  $R_m=-8\pi GT/c^4$ reads

\BEQ \label{RB=}  R_m&=&
-\frac{1}{\xi(1+B)}\left[ \frac{3\xi B''}{V}+\frac{6\xi rB'}{W^2}+\frac{B_0'{}^2}{4 B_0V}
\right.\\&+&\left.  2\mb^2B_0+\mb^2\lambdaB B_0^2
-\frac{2\bE_0^2B_0- \mu_\bi^2\xi B}{2 U}
+\frac{\xi\bar T_e}{\bM^2}\right].\nn
\EEQ
The last term is proportional to $ T_e=\rho_e- p_r^e-2 p_\perp^e$, 
the trace  of the energy momentum tensor from excited states.

The Gross-Pitaevskii equation, 

\BEQ \label{GP13}
\frac{\bE_0^2}{U}\psi_0+\frac{1}{V}\psi_0''+\frac{2r}{W^2}\psi_0'=(\mb^2+\xi R)\psi_0
+\frac{\lambda N}{2\hbar c}\psi_0^3,\EEQ

\ni  may be expressed in terms of $B_0\sim\psi_0^2$ and $B=B_0+B_e$, 


\BEQ \label{fullBeqn} && -6\xi B_0(\frac{B''}{V}+\frac{2rB'}{W^2}) 
-(1+B)(\frac{B_0''}{V}+\frac{2rB_0'}{W^2}) \nn\\&&
+\frac{(1+B_e)B_0'{}^2}{2B_0V}+2\bm^2B_0[(1+\lambdaB B_0)(1+B_e)-B_0] \nn\\&&
+\frac{2\bE_0^2- \mu_\bi^2\xi }{U}B_0^2
-\frac{ \mu_\bi^2\xi}{U}B_0B_e
=\frac{2\xi \bar T_e B_0}{\bM^2}. \EEQ


We can now first verify that the total energy momentum tensor is
conserved due to the harmonic condition (\ref{harm}). The terms
$B_0^2/U$ are singular because we shall consider $U(0)=0$ and $B_0(0)>0$; they drop out from 
(\ref{rhoB=},\ref{prB=},\ref{pperpB=}) 
and (\ref{RB=},\ref{fullBeqn}) for $\bE_0=\mu_\bi\sqrt{\xi/2}$, that is,

\BEQ \label{E02=} E_0^2= \half\xi\hbar^2c^2\mu_\bi^2=
\frac{8\pi G\hbar^2}{c^2}\xi\rho_\bi.
\EEQ

\ni We shall verify later that $B_e(0)=0$, so the $B_0B_e/U$ term is 
indeed less singular.
We may decompose $\xi$ as

\BEQ \xi=\xi_0\xi_1=\frac{\xi_0}{1+\xi_e}.\quad 
\EEQ
\ni We shall later see that $\xi_e$ is small. The leading term is 

\BEQ \xi_0\equiv \frac{2}{3}\bm^2\bM^2=\frac{2m^2M^2}{3m_P^4}=3.288\cdot 10^{55}M_9^2 
\EEQ

In the regime $B,B_0\gg1/\xi$, Eq. (\ref{fullBeqn}) simplifies, 

\BEQ \label{eqnBshortA}
&-&\frac{B_0''}{V}-\frac{2rB_0'}{W^2}+\frac{\bm^2}{3\xi}(1-B_0+\lambdaB B_0)=S_B^0\\
S_B^0&=&\frac{B_e''}{V}+\frac{2rB_e'}{W^2}-\frac{\bm^2}{3\xi}(1+\lambdaB B_0)B_e
+\frac{\mu_\bi^2B_e}{6U}+\frac{\bar T_e}{3\bM^2}, \nn
\EEQ

\ni In terms of $\psi_0$ the Gross-Pitaevskii equation (\ref{eqnBshortA}) reads

\BEQ \label{eqnBshortC}
-\frac{\psi_0''}{V}-\frac{2r\psi_0'}{W^2}-\frac{B_0'{}^2}{4B_0^2V}\psi_0&+&
\frac{\bm^2[1-(1-\lambdaB)B_0]}{6\xi B_0}\psi_0\nn\\
&=&\frac{ S_B^0}{2B_0}\psi_0.
\EEQ

\ni Either (\ref{eqnBshortA}) or (\ref{eqnBshortC}) determines $B_0$ or $\psi_0$,
once $B_e$ and $T_e$ are known. To leading order, they can be omitted. It is simpler, 
however, to work with $B=B_0+B_e$ rather than with $B_0$ itself, which satisfies

\BEQ \label{eqnBshort}
&&\frac{B''}{V}+\frac{2rB'}{W^2}-\frac{\bm^2}{3\xi}(1-B+\lambdaB B)=\frac{S_B}{2\bM^2},\\
S_B&=&\frac{2\bm^2\bM^2}{3\xi}[2-\lambdaB(1- B+B_e )]B_e
-\frac{\mub^2B_e}{3U}-\frac{2\bar T_e}{3}. \nn
\EEQ


\myskip{
It also holds exactly that

\BEQ
\label{m2+xiRm} 
\mb^2+\xi R_m 
+\frac{\lambda}{2\hbar c}N\psi_0^2=-\frac{B_0'{}^2}{4B_0^2V}+\frac{B_0''}{2B_0V}+\frac{rB_0'}{B_0W^2} .
\EEQ
}
\myskip{
\\&\approx  &  
-\lambdaB \mb^2 B-\frac{B'{}^2}{4B^2V}
+\frac{1-B+\lambdaB  B}{4B\bM^2}-\frac{\delta\bar T_m}{6B\bM^2}. \nn\EEQ

\ni where the first relation is exact and the second holds for $B\gg1/\xi$.
This expression allows to verify that the Gross-Pitaevskii equation 

\BEQ \label{GP13}
&&\frac{1}{V}\psi_0''+\frac{2r}{W^2}\psi_0'=(\mb^2+\xi R_m)\psi_0
+\frac{\lambda N}{2\hbar c}\psi_0^3=\\
&&\left (-\frac{B'{}^2}{4B^2V}-\lambdaB\mb^2B_e+\frac{1-(1-\lambdaB) B}{4B\Mb^2}-\frac{\delta \bar T_m}{6B\bM^2}\right)\psi_0,\nn
\EEQ

\ni coincides for $\psi_0\sim\sqrt{B_0}$ with (\ref{eqnBshort}), 
provided the excited states contribution $B_e$ to $B$, 
see (\ref{Bdef}),  is  neglected.}

It also holds that

\BEQ
\label{m2+xiRm} 
\mb^2 \!\!&\!\!+\!\!&\!\! \xi R_m= 
-\lambdaB \mb^2 B-\frac{B'{}^2}{4B^2V}+\frac{B''}{2BV}+\frac{rB'}{BW^2} 
\\&\approx  &  
-\lambdaB \mb^2 B-\frac{B'{}^2}{4B^2V}
+\frac{1-B+\lambdaB  B}{4B\bM^2}-\frac{\delta\bar T_m}{6B\bM^2}. \nn\EEQ

\ni where the first relation is exact and the second holds for $B\gg1/\xi$.

\section{Exact solution in the interior}

Since $\lambdaB$ will turn out to be very small, we first take it zero.
We also neglect the excitation terms $B_e$ and $T_e$.
Solving Eq. (\ref{eqnBshort}) for $\lambdaB=0$ and a constant $B$, we
find a relation and, with (\ref{E02=}) as consequence,

\BEQ \label{Psiconds} B=1, \quad
N|\psi_0^2|=\frac{E_0}{8\pi\xi}=\frac{\rho_\bi}{E_0}=\frac{\mu_\bi}{8\pi\sqrt{2\xi}\ell_P^2}. \EEQ

\ni 
The value $B=1$
expresses a $100\%$ direct backreaction of matter on the metric.

 Instead of searching a finite $U$, as for boson stars, ~\cite{bosonstar1,bosonstar2} 
we assume a very small $U$ with $U(0)=0$, coded by a parameter $u$, of the form

\BEQ \label{U=}
 U=\frac{1}{2}u\mu_\bi^2W^2
\EEQ

\ni It is custumary to introduce the mass function ${\rm M}(r)$ and 
$\M(r)=G{\rm M}(r)/c^2$, defined by

\BEQ \label{VMdef} V=\frac{W'{}^2}{1-2\M/W}.\EEQ

\ni The $00$ and $11$ Einstein equations then take the form

\BEQ \label{M'=}
&&\M'=\frac{4\pi G}{c^4} W' W^2\rho_{\rm tot},\quad \!\! \nn\\
&&(\half W-\M)\frac{U'}{U}-\frac{\M W'}{W}=
\frac{4\pi G}{c^4} W^2W'\,p^r_{\rm tot}.
\EEQ

\ni We may combine (\ref{rhotot=}) and (\ref{rhoB=}) together with $B=1$
and the Ansatz (\ref{U=}), to obtain

\BEQ\label{rptotout} (\rho_{\rm tot},p_r^{\rm tot})=\frac{c^4}{4\pi G W^2}
\left(\frac{1}{4 u }\pm\frac{\mb^2W^2}{8\xi }\right). 
\EEQ


\ni In RTG there is a solution of (\ref{M'=}) and (\ref{rptotout}),
\BEQ \label{Einout}
\M=\frac{W}{4}+\frac{\mb^2W^3}{24\xi},\quad u=1.
\EEQ   

\ni It would not exist within GTR, as is seen by taking $\mu_\bi\to0$,
$u\to\infty$ first.
For the Schwarzschild black hole the horizon occurs when $\M=\bM$ for $W=2\bM$.
Concerning the outside metric, we will be close to that situation.
This implies that a mass $M$ requires the large, non-constant coupling $\xi=\xi_0\xi_1$ with

\BEQ \xi_0\equiv \frac{2m^2M^2}{3m_P^4}=3.288\cdot 10^{55} \left(\frac{M}{10^9M_\odot}\right)^2.
\EEQ

\ni and $\xi_1=1$ up to possible small corrections to be discussed further on. 
Interestingly, $\xi_0$ is a measure for the area of the black hole ($4\pi \bM^2$) 
expressed in units of the square of the hydrogen Compton length $1/\bm=\hbar/mc$,
so it is a measure of the surface entropy.  Notice that in the
Bekenstein-Hawking entropy the area is expressed in squares of the Planck length.
In this work we shall consider the $T=0$ situation, so that the entropy is exactly zero.

We now get

\BEQ E_0=\mu_\bi\sqrt{\half\xi}=\frac{1}{\sqrt{3}}\mu_\bi GMm=\frac{1}{\sqrt{3}}\mub mc^2.
\EEQ

\ni and coupling

\BEQ 
\label{lambdaB=}
\lambdaB &=& \frac{\lambda m_P^2}{32\pi m^2\xi} =
\frac{3a_s\hbar^2c^4}{2G^3m^3M^2}=4.166\cdot10^{-13}
\frac{1}{M_9^2}. \nn
\EEQ

\ni Though the latter is small, the product $\lambdaB\xi$ is large,

\BEQ \lambdaBB\equiv 6\lambdaB \xi=\frac{6a_sc^2}{Gm}=8.19\cdot 10^{43}.
\EEQ

\myskip{

\BEQ \lambda&=&\frac{8gE_0^2}{\hbar^3c^3}
=\frac{32\pi}{3}\mub^2 a_s\bm=1.527\cdot10^{-21}M_9^2,
\\ 
\label{lambdaB=}
\lambdaB &=& \frac{\lambda m_P^2}{32\pi m^2\xi} =
\frac{a_s\hbar^2\mu_\bi^2}{2Gm^3}=7.825\cdot 10^{-41}.\EEQ

\ni Though the latter is extremely small, $\lambdaB\xi$ is large,

\BEQ \lambdaBB\equiv 6\lambdaB \xi=\frac{2\mub^2a_sc^2}{Gm}=
1.54\cdot 10^{16} \left(\frac{M}{10^9M_\odot}\right)^2.
\EEQ
}

Let us introduce the `Riemann' variables $x$ and $y$ by 

\BEQ
x=\frac{W}{2\bM},\qquad y=\sqrt{1-x^2},\qquad \EEQ

\ni so that
$U=2\mub^2x^2$. With (\ref{U=}), (\ref{VMdef}) and $\M=\half
\bM(x+x^3)$ from (\ref{Einout}), the harmonic constraint
(\ref{harm}) brings

\BEQ \frac{2x'}{x}-\frac{2x''}{x'}-\frac{2xx'}{1-x^2}
+\frac{4x'}{x}=\frac{8rx'{}^2}{x^2(1-x^2)}. \nn \EEQ

\ni Going to the inverse function $r(x)$ makes it linear,

\BEQ\label{rxdeqn} x^2(1-x^2)r_{xx}+x(3-4x^2)r_x=4r. \EEQ

\ni In terms of the variable $y$ this transforms into

\BEQ x^4r_{yy}-4x^2yr_y=4r. \EEQ

\ni The solution is then remarkably simple,

\BEQ\label{rx=}
r=\frac{r_1}{\sqrt{1-y^2}}(1+\frac{y}{\sqrt{5}})
\left(\frac{1-y}{1+y}\right)^{\sqrt{5}/2}.
\EEQ

\ni (The second independent solution with $\sqrt{5}\to-\sqrt{5}$ or $y\to-y$
is singular at $r=0$, $y=1$.) It will hold that 
$r_1\approx\bM$. This determines the metric functions $W'$ and $V$, 

\BEQ
\label{W'x=}  W'&=&\frac{\sqrt{5}}{2}x^{2-\sqrt{5}}y(1+y)^{\sqrt{5}},\nn\\
\qquad \label{Vx=}
V&=&\frac{2W'{}^2}{y^2}=\frac{5}{2}x^{4-2\sqrt{5}}(1+y)^{2\sqrt{5}}.
\EEQ

Putting these results together, it now follows that
\BEQ
 \rho=\frac{3c^4}{64\pi G\bM^2} ,\quad  p_r=p_\perp=p\equiv-\frac{3c^4}{64\pi G\bM^2},
\EEQ
So in the interior we reproduce the vacuum equation of state $\rho=-p=$ const.,
that is,  $\bar\rho=-\bar p=3/8$.

To understand the structure of the problem, we again take
$\lambdaB=0$. Then for any $A$ there is the solution 

\BEQ
\label{BA=} 
B(x)=1+Ay=1+A\sqrt{1-x^2}.
\EEQ
 
\ni Surprisingly,  their $A$-dependence factors out, keeping a vacuum
equation of state $\bar\rho=- \bar p=3/8$, so (\ref{BA=}) is an exact,
non-uniform solution of the same metric. It can be verified  that
 $\psi_0\sim\sqrt{1+Ay}$ solves the Gross-Pitaevskii equation 
 (\ref{eqnBshort}) at $\lambdaB=0$ and without excited states terms, as it should.

\subsection{Normalization}

To normalize $\psi_0$, we need
the $3d$ volume element in the future time direction,
$\d \Sigma^\mu= \d r\d\theta\d\phi n^\mu \sqrt{-g_3}\equiv \delta^\mu_0\d {\rm V}$,
set by the timelike unit vector $n^\mu=\delta^\mu_0/\sqrt{U}$
and $g_3=-VW^4\sin^2\theta$. This results in
$\d {\rm V}=\d r \d\Omega \sqrt{V/U\,}\,W^2$. 
It then holds that $\sqrt{-g}\d^4x=U\d x^0\d {\rm V}$.

The general inner product ~\cite{BirrellDavies}
$(\psi_1,\psi_2)=(-i/\hbar c)\int \d \Sigma^\mu$
$(\psi_1\p_\mu\psi_2^\ast-\p_\mu\psi_1\psi_2^\ast)$ defines the
orthonormality, already noticed below (\ref{orthonorm}),

\BEQ \label{innerprod}
(\psi_i,\psi_j)=\frac{E_i+E_j}{\hbar^2c^2}\int \d{\rm V}
\,\psi_i\psi_j^\ast\equiv\delta_{ij}, \EEQ 

\ni together with $(\psi_i,\psi_j^\ast)=0$
and $(\psi_i^\ast,\psi_j^\ast)=-\delta_{ij}$.
With the volume element

\BEQ \d{\rm V}=\d y\d\Omega\,8\bM^2/\mu_\bi\EEQ

\ni it yields a volume ${\rm V}=32\pi\bM^2/\mu_\bi=4\pi(2\bM)^3/\mub$ and

\BEQ \label{Psi02=}
|\Psi_0^2| =\frac{2E_0}{\hbar^2c^2}N_0|\psi_0^2|=\frac{\sqrt{3}\mub c^6}{16\pi G^3mM^2}(1+Ay).
\EEQ

\subsection{Properties of the solution}

The horizon is located at $r=\bM$, $x=1$, $y=0$.
We may integrate (\ref{Psi02=}) over the BH, which yields $N_0$.
Alternatively, from the definition (\ref{Bdef}) of $B$, we may consider $\int\d{\rm V}\,B$, 
making use of the normalization $(\psi_0,\psi_0)=1$. Either way, this yields

\BEQ N=N_0+N_e=2\sqrt{3}\frac{M}{m}\int_0^1\d y\,[B_0(y)+B_e(y)].
\EEQ

\ni It allows to relate the BH energy to the groundstate occupation,

\BEQ\label{MNnu=}
 Mc^2=\nu N_0mc^2,\qquad \nu=\frac{1}{(2+A)\sqrt{3}}. \EEQ

\ni Clearly, the energy $Mc^2$ of the BH can be at best $29\%$  of the rest energy
$N_0mc^2$ of the constituent hydrogen atoms. In the BH formation, the major part of
the energy, $(1-\nu)N_0mc^2$ minus the potential energy, 
has to be radiated out. This may explain the large luminosity of quasars.

The leading part of the energy of the quantum field,

\BEQ
E_\psi^{(0)}\equiv\int\d{\rm V} \frac{\lambda}{4\hbar c}\langle\hat\psi^\dagger{}^2\hat\psi^2\rangle=
\frac{3\lambdaB}{2 \mub} Mc^2\int_0^1\d y\,B^2
\EEQ

\ni equals 

\BEQ E^{(0)}_\psi 
=\frac{9a_s\hbar^2c^8}{4\mu_\bi G^4m^3M^2}(1+A+\frac{A^2}{3}).
\EEQ

\ni With $\lambdaB/\mub=17.5/M_9^3$ it is of the order of  the total energy $Mc^2$,
and it may exceed it. To obtain the total energy, the gravitational energy density has to 
be taken into account, which is partly positive and partly negative. Due to a sum rule the total
energy is always $Mc^2$~\cite{NEPLBH08}.


At the origin the solution exhibits the powerlaw singularities 

\BEQ U=\bar
U_1r^{\gamma_\mu},\quad V=\half\gamma_\mu^2\bar
W_1^2r^{\gamma_\mu-2},\quad W=\bar W_1r^{\half\gamma_\mu},\EEQ

\ni where $\gamma_\mu=\half(\sqrt{5}+1)$ is the golden mean. But if we
take $W$ as the coordinate, we have in the interior the  shape

\BEQ \d{\rm s}^2=\half\mu_\bi^2W^2c^2\d t^2-\frac{2\d W^2}{1-{W^2}/{4M^2}}-
W^2\d\Omega^2, \EEQ 

\ni which is regular at its origin, with the term
$2\d W^2$ coding the above powerlaw singularities in $r$.

\section{General solution near the horizon}

Near the horizon the exact solution will be deformed.
In general we may code the functions $U(r)$ and $V(r)$
in new functions $u(r)$ and $v(r)$, 

\BEQ U=2\mub^2x^2u, \quad V=\frac{8\bM^2x'{}^2}{y^2}v,\EEQ
and $W=2\bM x$. With (\ref{harm}) we then have for arbitrary $f$

\BEQ \frac{f_{rr}}{V}+\frac{2rf_r}{W^2}=\frac{x^2f_{yy}}{8\bM^2v}+
[-4y+\frac{x^2}{2}(\frac{u_y}{u}-\frac{v_y}{v})]\frac{f_y}{8\bM^2v},
\nn
\EEQ

\ni since the $W''$ terms cancel, as they should, because one can also start with $W$ 
as variable instead of $r$. 
The  leading shape of the Gross Pitaevskii Eq. (\ref{eqnBshort}) can be written as 
function of $y$,
(we neglect $\lambdaB$, as it will be much smaller than $B_e$ and $\bar T_e$)

\BEQ \label{SBedef}
&&
\frac{x^2}{4v}B_{yy}-
\frac{y}{v}B_y+\frac{x^2}{8v}(\frac{u_y}{u}-\frac{v_y}{v})B_y+B-1=S_B^e\nn\\
&&S_B^e=2(1+\xi_e)B_e-\frac{1}{6x^2}B_e -\frac{2}{3} \bar T_e-\xi_e(B-1) .
 \EEQ

\ni In this equation $B=B_0+B_e$ consists of contributions of both the groundstate and 
excited states, while $B_e$ and $\bar T_e$ involve the latter only. While $v=1$ in the BH
interior,  it grows as $\sim e^\zeta/\eta$ beyond the horizon, possibly enhancing the effect of 
the flucutations. In principle this equation may therefore
describe a decay of $B_0$ to zero, embedded in excited states that decay slower.

The Einstein equations read in terms of the new functions $u$ and $v$

\BEQ \frac{1}{x^2}(2-\frac{y^2}{v})+\frac{1}{v}(2-\frac{yv_y}{v})
&=&\frac{1}{x^2u}+8\bar\rho ,    \\
\frac{2}{x^2}-\frac{3y^2}{x^2v}+\frac{yu_y}{uv}&=&-\frac{1}{x^2u}-
8\bar p_r. \nn
\EEQ

Expressing the shapes (\ref{prB=}) in $y$, we have

\BEQ
\bar \rho&=&
 \frac{{x^2}B_{yy}-(3y+\frac{x^2v_y}{2v})B_y}{8v(1+B)}+\frac{6B_0
 +3\lambdaB B_0^2}{8\xi_1(1+B)}+\frac{\bar\rho_e}{1+B},\nn\\
\label{rhopBx}  
\bar p_r&=& \frac{(3y-\frac{x^2u_y}{2u})B_y}{8v(1+B)}
-\frac{6B+3\lambdaB B_0^2}{8(1+B)}
+\tilde p_r^e,
\\
\bar p_\perp&=&
 \frac{-x^2B_{yy}+[3y-\frac{x^2}{2}(\frac{u_y}{u}-\frac{v_y}{v})]B_y}{8v(1+B)}
 -\frac{6B_0+3\lambdaB B_0^2}{8\xi_1(1+B)}\nn\\&&
+\frac{\bar p_\perp^e}{1+B},\nn
 \EEQ

\ni In the region $B\gg1/\xi$ we have the simplifications

\BEQ \label{rhoesimpl}
 \bar \rho &=& \frac{(y-\frac{x^2u_y}{2u})B_y}{8v(1+B)}
 +\frac{4+2B+\lambdaB(4B+3 B^2)}{8(1+B)}+\tilde\rho_e,
\nn\\
\bar p_\perp&=& 
 \frac{-yB_y}{8v(1+B)}
 -\frac{4+2B+\lambdaB(4B+3 B^2)}{8\xi_1(1+B)}\nn\\&&+
\frac{\bar p_\perp^e-(\frac{1}{\xi_1}-\frac{1}{12x^2})B_e+\frac{1}{3}\bar T_e}{1+B}.\nn
\EEQ

\ni with the following source terms at $\lambdaB=0$

\BEQ \tilde\rho_e&=& \frac{\xi_e(2+B)+
4\bar\rho_e-3(1+\xi_e)B_e+2S_B^e}{4(1+B)},\nn\\
\tilde p_r^e&=&-\xi_e\frac{6B 
}{8(1+B)}+\frac{4\bar p_r^e+3(1+\xi_e)B_e}{4(1+B)}.
\EEQ

The equation (\ref{rxdeqn}) for the Minkowski coordinate now reads

\BEQ\label{rxdeqnu} x^2y^2r_{xx}+x\left[4y^2-1
+\frac{xy^2}{2}(\frac{u_x}{u}-\frac{v_x}{v})\right]r_x=4vr. \EEQ

\section{Excitations}


Let us reformulate our theory on a new basis.
We go to a new coordinate $z$, 
\BEQ x=\frac{W}{2\bM}=\frac{1}{\cosh (z/\sqrt{2})},\quad 
y=-\tanh \frac{z}{\sqrt{2}},\EEQ

\ni so that $z=-\infty$ at $r=0$ and $z=0$ at $r=\bM$,  a dimensionless time $s$
and a scaled energy,

\BEQ s=\frac{\mu_\bi c t}{\sqrt{2}},\quad \tilde E=\frac{\sqrt{2}}{\hbar c\mu_\bi}E.  \EEQ

\ni The line element then becomes 

\BEQ \d {\rm s}^2=4\bM^2x^2(u\d{\rm s}^2-v\d z^2-\d\Omega^2),\EEQ

\ni which is Minkovskian in the exact solution where $u=v=1$.
It corresponds to volume elements

\BEQ && \d^4r\sqrt{-g}=
\d^4\tilde r\sqrt{uv}\,16 \bM^4 x^4,\quad \d^4\tilde r=\d s\d z\d\Omega\,\\&& 
\d{\rm V}=\d y\d\Omega\,\frac{8\bM^2}{\mu_\bi}\sqrt{\frac{v}{u}}
=\d z\d\Omega\,\frac{4\bM^2x^2}{\mu_\bi}\sqrt{\frac{2v}{u}}
\EEQ

\ni In terms of the new field 

 \BEQ 
 \del \hat\psi\equiv\hat\psi-\hat a_0\psi_0 =\frac{\sqrt{\hbar c}\, v^{1/4}\hat\chi}{2\bM u^{1/4}x} \EEQ

\ni the innerproduct $(\del\hat\psi_i,\del\hat\psi_j)$ just becomes

\BEQ (\chi_i,\chi_j)\equiv (\tilde E_i+\tilde E_j)\int\d z\d\Omega\frac{v}{u} \chi_i^\ast\chi_j
=\delta_{ij}.\EEQ

\ni For the kinetic term it holds that

\BEQ&&\d^4r\sqrt{-g} \p^\mu\delta\hat\psi^\dagger\p_\mu\delta\hat\psi=
\d^4 \tilde r\,v\hbar c 
\left(\frac{\p_s\hat\chi^\dagger \p_s \hat\chi}{u} \right.\\
&-&\left. \frac{u^{1/2}x^2}{v^{3/2}}(\hat\chi^\dagger_1)_z(\hat\chi_1)_z
+{\bf L}\hat\chi^\dagger\cdot {\bf L}\hat\chi\right)
\nn
\EEQ

\ni where $\hat\chi_i=(v/u)^{1/4}\hat \chi/x$ and ${\bf L}=-i[\p_\theta,(1/\sin\theta)\p_\phi]$ 
is the angular momentum operator in units of $\hbar$. 
Likewise, the $-(\bm^2+\xi R)\delta\hat\psi^\dagger\delta\hat\psi$ term 
becomes

\BEQ \label{B3=} (6\lambdaB\xi_0 Bx^2-V_{01}-\frac{\xi}{u})
\frac{\hbar c\sqrt{v}\,\hat\chi^\dagger\hat\chi}{16\bM^4x^4\sqrt{u}},\EEQ

\ni with potential

\BEQ
V_{01}=-\frac{x^4B_y^2}{8vB^2}-(1-\lambdaB-\frac{1}{B})x^2
-\frac{2x^2\bar T_e}{3B}.\nn\EEQ

\ni In vacuum, $z\to 0^+$, $R_m\to0$ and $V_{01}\to6\xi$, which 
acts as an infinite barrier. More precisely, it leads to a decay
at a scale of the Compton wavelength $\hbar/mc=1/\mb$.

\ni Finally the interaction term brings quadratic terms from the expansion 
$\hat\psi=\hat a_0\psi_0+\delta \hat\psi$
up to second order. Noting the time-dependence $\psi_0=|\psi_0|e^{-i\tilde\mu s}$ with
\BEQ\tilde\mu=\tilde E_0=\sqrt{\xi},\EEQ

\ni  this yields in a straightforward manner

\BEQ \label{B4=} -\frac{3\lambdaB\xi_0
 B_0 \hbar c\sqrt{v}}{16\bM^4x^2\sqrt{u}}(4\hat\chi^\dagger\hat\chi+e^{2i\tilde\mu s}
\hat\chi^2 +e^{-2i\tilde\mu s}\hat\chi^\dagger{}^2), \EEQ

\ni where $B_0$ is the groundstate contribution to $B$. 
Notice that (\ref{B3=}) and (\ref{B4=}) involve the same prefactor 
$\lambdaB\xi_0=\lambdaBB/6\xi_1$. Together this brings an action

\BEQ S_{\rm mat}=\hbar\int\d s\d z\d\Omega\,v(L_2+L_{\rm int})\EEQ

\ni with, after a partial integration, a Lagrangian density

\BEQ
\label{L2=} &L_2&= \frac{\p_s\hat\chi^\dagger\p_s\hat\chi}{u} -
\frac{1}{v}\p_z\hat\chi^\dagger\p_z\hat\chi
+{\bf L}\hat\chi^\dagger\cdot {\bf L}\hat\chi \\&&
-(\frac{\xi}{u}+V_{0}+V_1)\hat\chi^\dagger\hat\chi 
-\half V_1(e^{2i\tilde\mu s}\hat\chi^2
+e^{-2i\tilde\mu s}\hat\chi^\dagger{}^2),  \nn \EEQ

\ni which involves the potentials

\BEQ
V_{0}(z)&=&-\frac{x^4B_y^2}{8vB^2}-(1-\lambdaB-\frac{1}{B})x^2
-\frac{2x^2\bar T_e}{3B}\nn\\
&+&\frac{1-2x^2}{2v}+\frac{y}{2\sqrt{2}}(\frac{u'}{u}-\frac{v'}{v}) \\
&+&\frac{1}{16v}\left[
(\frac{4u''}{u}-\frac{4v''}{v})-(\frac{u'}{u}-\frac{v'}{v})(\frac{3u'}{u}+\frac{5v'}{v})\right],
\nn\\
 V_1(z)&=&\frac{6 \lambdaB\xi_0 B(z)}{\cosh^2(z/\sqrt{2})}=\frac{\lambdaBB}{\xi_1}Bx^2.\nn\EEQ

The equation of motion for the $\hat\chi$ field is

\BEQ -\frac{\p_s^2\hat\chi}{u}= -\frac{\p_z^2\hat\chi}{v}+(\frac{\xi}{u}+L^2+V_{0}+V_1)\hat\chi+
V_1e^{-2i\tilde\mu s}\hat\chi^\dagger.\nn
\EEQ


\ni We have to perform a Bogoliubov transformation in our case with non-constant 
``potential'' $\xi R$, a situation similar to the case of non-constant external 
potential, discussed in e. g. ~\cite{PitaevskiiStringari}.
With $i\equiv(n,\ell,m)$ we set 
\footnote{The complex Bogoliubov functions $u_i$ and $v_i$ should not
be mistaken for the positive metric functions $u$ and $v$.}

\BEQ \hat\chi(z,\theta,\phi)=&\sum_i&\left[
\,u_i(z)Y_{\ell m}(\theta,\phi)e^{-i(\tilde\mu+\omega_i) s}\hat b_i
\right. \\&&+\left. v_i^\ast(z)  Y^\ast_{\ell m}(\theta,\phi)
e^{-i(\tilde\mu-\omega_i)s}\hat b_i^\dagger \right],   \nn
\EEQ

\ni and the conjugate momentum is $(v/u)\p_s\hat\chi$. 
With $\tilde {\bf r}=(z,\theta,\phi)$ the commutation relation (\ref{comrel=})
becomes

\BEQ 
&&\frac{v}{u}\sum_i (\tilde\mu+\omega_i)[ u_i(\tilde {\bf r})u_i^\ast(\tilde {\bf r}')
+u_i^\ast(\tilde {\bf r})u_i(\tilde {\bf r}')] \\
&-&\frac{v}{u}\sum_i(\tilde\mu-\omega_i)[
v_i(\tilde {\bf r})v_i^\ast(\tilde {\bf r}')+v_i^\ast(\tilde {\bf r})v_i(\tilde {\bf r}')]
=\delta(\tilde{\bf r}-\tilde{\bf r}')\nn.
\EEQ

\ni where $u_i(\tilde {\bf r})=u_i(z)Y_{\ell m}(\theta,\phi)$.
We may then expect the orthonormality property


\BEQ  \label{inprodnew}
\int\d z\,\frac{v}{u}\,
[ u_i^\ast(z)u_{j}(z)-v_i^\ast(z)v_{j}(z)]=\frac{\delta_{ij}}{2\tilde\mu}.
\EEQ

\ni The eigenmodes satisfy the Popov equations

\BEQ \frac{2\tilde\mu\omega_i}{u}u_i&=&-\frac{u_i''}{v}
+\left[V_0+\ell(\ell+1)-\frac{\omega_i^2}{u}\right]u_i\nn\\&&+V_1(u_i+v_i),\nn\\
-\frac{2\tilde\mu\omega_i}{u}v_i&=&-\frac{v_i''}{v}
+\left[V_0+\ell(\ell+1)-\frac{\omega_i^2}{u}\right]v_i\nn\\&&+V_1(u_i+v_i),
\EEQ

\ni where derivatives are with respect to $z$.
 From these equations one can show that

\BEQ (\omega_i-\omega_j^\ast)\int\d z\frac{v}{u}(u_iu_j^\ast-v_iv_j^\ast)=0\EEQ

\ni and with real $\omega_i$ this confirms the inner product 
(\ref{inprodnew}). 

In the interior one has $u=v=1$.
At  $\omega_i=\ell=0$ and $\bar T_e\to0$ the above equations then allow the exact solution
 $u_0=-v_0=x\sqrt{1+Ay}$,  which corresponds to $\pm \psi_0$ on this basis,
so the solvability stems from the one of the Gross-Pitaevskii equation.
This situation is related a gauge transformation 
that changes the phase of the groundstate  $\psi_0$~\cite{PitaevskiiStringari}.

For excited states $\omega_i\mu\sim\ell^2\sim V\sim\lambdaBB$  
are large, so both $V_0$ and $\omega_i^2$  can be neglected.
Then $\omega\sim\lambdaB\tilde\mu$. 
In the regime $z\ll-1$ one has $V\ll1$, so one expects

\BEQ u_i= c_n e^{ik_nz},\quad v_i=-\frac{c_n}{2k_n^2}V e^{ik_nz}.
\EEQ

\ni We can deal with the boundary conditions of these excited states as with 
plane waves, e. g. by requiring
that $e^{ik_nz}=1$ at some large $z=-L$ and take $L\to\infty$ at the end.
The normalization constant is then

\BEQ 
c_n=\frac{1}{\sqrt{2\tilde\mu L}}
\EEQ
 
Another regularization is to assume that $k=k'-ik''$ has a small imaginary part;
this will keep all integrals starting at $-\infty$ finite. Then 
$c_n\sim\sqrt{k''}$. Below we shall employ sine-modes and impose the hard wall 
boundary condition at $z=-L$.

\subsection{Between the center and the peak of the potential}

In the typical case where $A>0$  the potential $V=\lambdaBB(1-y^2)(1+Ay)$ 
has a maximum at

\BEQ z_c=-\sqrt{2}\,{\rm arctanh} \,y_c,\quad y_c=\frac{\sqrt{1+3A^2}-1}{3A},\EEQ

\ni which goes to zero for $A\to0$, but remains finite for $A\to\infty$.
The region $-\infty<z<z_c$, which covers the whole interior when $A\to0$,
is considered first.

Since $V=\lambdaBB x^2B$ is large, the excited states can be analyzed with the WKB method.
The function

\BEQ s=u_i+v_i,\quad \EEQ
satisfies with

\BEQ \omega_i\equiv  \frac{k^2}{2\sqrt{\xi}},\qquad \ell(\ell+1)\approx \ell^2\EEQ

\ni the equation 

\BEQ k^4s=s^{iv}-2\ell^2s''+\ell^4s+\ell^22Vs-2(Vs)''. 
\EEQ

\ni We make the Ansatz

\BEQ s(z)={\rm const.}\,e^{iS(z)-\tau(z)}\EEQ

\ni where $S=O({\lambdaBB}^{1/2})$, $\tau=O(\lambdaBB^0)$ and higher order corrections
may be neglected. The role of $L$ will be discussed below. We have at leading order

\BEQ S'{}^4+2(V+\bar\ell^2)S'{}^2+\ell^4+2V\ell^2=k^4.
\EEQ

\ni with the solution

\BEQ \label{S'=}
S'{}^2=\sqrt{k^4+V^2}-V-\ell^2.\EEQ

\ni At a given location $z$ the solution is of plane wave type when $S'{}$ is real,
which occurs provided $\ell$ is limited,

\BEQ \ell^2\le \ell_+^2(z)\equiv \sqrt{k^4+V^2(z)}-V(z)
\EEQ
  
\ni We shall not need $S$ itself. At next order we find 

\BEQ 4(S'{}^3-S'V-S'\ell^2)\tau'=2(S'{}^3-S'V-S'\ell^2)'.\EEQ

\ni with solution fixed to $\tau=0$ for $z\to-\infty$ ($V\to0$),

\BEQ \tau=\frac{1}{4}\ln\frac{k^4+V^2}{k^4}
+\frac{1}{4}\ln\frac{\sqrt{k^4+V^4}-V-\ell^2}{k^2-\ell^2},
\EEQ

\ni  To get $u_i$ and $v_i$ to leading order is now easy. Since

\BEQ u_i-v_i=\frac{(\ell^2+2V)s-s''}{k^2}=\frac{V+\sqrt{k^4+V^2}}{k^2}s,\EEQ

\ni and imposing  $S(z_i)=0$, we end up with

\BEQ \label{ui=vi=}
(u_i,v_i)&=&\frac{1}{\sqrt{\tilde\mu L}}\,\frac{k^2\pm(V+\sqrt{k^4+V^2})}{2k(k^4+V^2)^{1/4}}
\nn\\&\times&
\left(\frac{k^2-\ell^2}{\sqrt{k^4+V^2}-V-\ell^2}\right)^{1/4}\,\sin S.
\EEQ

\ni For $z\to-\infty$ ($V\to0$) one has indeed 
$u_i\to \sin kz/\sqrt{\tilde\mu L}$, $v_i\to0$. If we impose a hard wall boundary
condition $u_i=0$, $v_i=0$ at $z=-L$, the normalization (\ref{inprodnew}) is 
satisfied, since it is determined by values $z\ll-1$.
As costumary for plane wave problems, $L$ will be taken to infinity at the end.

For $\bar\ell>\bar\ell_+(z)$, the action $S$ becomes imaginary,
expressing a damping of the wave that has to penetrate the potential barrier 
to reach this position $z$. These states lead to negligible corrections.

We can calculate the Hamiltonian
\BEQ 
 \hat H={\rm const.}+\sum_i\tilde E_i^2\hat b_i^\dagger\hat b_i,\EEQ

\ni with

\BEQ \tilde E_i^2&=&\int\d z [\,(|u_i'|^2+|v_i'|^2+
\ell^2+\xi+V)(|u_i|^2+|v_i|^2)\nn \\&+&V(u_iv_i^\ast+u_i^\ast v_i)].
\EEQ

\ni With $|u_i'|^2+|v_i'|^2=S'{}^2(|u_i|^2+|v_i|^2)$,  this becomes

\BEQ 
\tilde E_i^2 &=&
\int\d z[(\xi+\sqrt{k^4+V^2})(|u_i|^2+|v_i|^2)\nn\\&+&V(u_iv_i^\ast+u_i^\ast v_i)].
\EEQ

\ni As in the normalization, these integrals are dominated by large negative $z$-values, 
where $V\to0$, so that we simply get

\BEQ \tilde E_i=\sqrt{\xi+k^2}\approx \tilde\mu+\frac{k^2}{2\tilde\mu}.
\EEQ

\ni It is degenerate (independent of $\ell,m$) as in the quantum Hall effect, 
though here there are no spectral gaps. At a given location $z$ the wavefunctions 
$u_i,v_i$ are oscillating (not damped) provided $\ell\le \ell_+(z)$.

\subsubsection{Contribution to the fraction of excited states}


We can now consider  $B_e$,  the excited states contribution to the direct back reaction 
$B=B_0+B_e$, that arises due to the interaction term, even at $T=0$.
We start from the definition

\BEQ B_e=\frac{16\pi G}{c^4}\xi\langle\delta\hat\psi^\dagger\delta\hat\psi\rangle\EEQ

\ni and express this as

\BEQ \label{Be=epsbBe}
B_e&=&
\frac{8\pi m^2\sqrt{v}}{3m_P^2x^2\sqrt{u}}\langle\hat\chi^\dagger\hat\chi\rangle
\equiv \varepsilon 
\bar B_e(z),\nn\\ 
\varepsilon&=&\frac{4\pi m^2\lambdaBB^{3/2}}{3m_P^2\sqrt{\xi}},\quad 
\bar B_e=\frac{2\sqrt{\xi v}}{\lambdaBB^{3/2}x^2\sqrt{u}}\sum_i |v_i|^2.
\EEQ

In the exactly solvable case
the sine modes, that exist for $z\le z_c$, i. e., to the left of the peak of $V$,
behave as $\sin kz$ for $z\to-\infty$. The hard wall boundary condition
$u_i=0$, $v_i=0$ at $z=L$ brings the quantization $k_n=n\pi/L$, $n=1,2,\cdots$.
Replacing $\sin^2S\to\half$, this yields at a given position $z$, 

\BEQ 
\bar B_e&=&\frac{2\sqrt{\xi}}{\lambdaBB^{3/2}x^2}\frac{L}{\pi}\int_0^\infty
\d k\int_0^{\ell_+^2(z)}\d\ell^2|v_i|^2 \nn \\
&=&\int_0^\infty\frac{\d k}{\pi}\frac{(K^2+V-k^2)^2}{4\lambdaBB^{3/2}x^2k^2K^2}\left[\,k\sqrt{K^2-V}\right. \\
&+&\left.(k^2+V-K^2){\rm arcsinh} \sqrt\frac{K^2-V}{k^2+V-K^2}\right].\nn
\EEQ

\ni 
where $V\equiv V(z)$ and $K^4\equiv k^4+V^2$. The integral gives

\BEQ \bar B_e(z)=\frac{0.236792}{x^2\lambdaBB^{3/2}}\,V^{3/2}=0.236792\,xB^{3/2}.
\EEQ

\ni so the result neatly vanishes at the origin ($x\to0$),

\BEQ B_e=0.757\,B^{3/2}x\,\frac{M}{10^9M_\odot},
\EEQ

\ni   According to (\ref{Be=epsbBe}) is has the characteristic strength

\BEQ \label{varep}
\varepsilon=\frac{4\pi m^2\lambdaBB^{3/2}}{3m_P^2\sqrt{\xi}}=
\frac{24\pi a_s^{3/2}c^2}{G^{3/2}m^{1/2}M}=0.382\frac{10^9M_\odot}{M}.
\EEQ

In standard BEC, the correction to the groundstate energy is
of relative order $\sqrt{na_s^3}$~\cite{PitaevskiiStringari}. 
Using (\ref{MNnu=}) we find in the homogeneous case $A=0$

\BEQ \sqrt{n_Ha_s^3}=0.0136\,\frac{10^9M_\odot}{M}, 
\EEQ

\ni so our variable $\varepsilon$ is of the same order of magnitude,
confirming the expectation of the introduction that the relevant physical
parameter is $na_s^3$, which is small for $M\gg10^9M_\odot$.

\subsection{Between the peak of the potential and the horizon}

We take $V=\infty$ beyond the horizon. Then it has quasi-bound states in the region 
$z_c<z<0$; they are not true bound states because
$V$ drops to zero for $z<<z_c$. But since the energy barrier $\sim \lambdaBB$ 
will be very large, the tunneling into the interior will be extremely small, and we shall neglect it.

For $U=V=1$ we can now copy previous solution (\ref{S'=}), (\ref{ui=vi=}). 
At given value of $\ell$,  a real valued $S'$ starts at $z=z_i$ set by

\BEQ V(z_i)=\frac{k^4-\ell^4}{2\ell^2} .\EEQ

\ni The surface state ``lives'' in the interval $z_i\le z\le0$.
The smallest $z_i$ arises when $z_i=z_c$ and it has $\ell_c^2=\sqrt{k^4+V_c^2}-V_c$,
while $z_i\to0$ for the maximum $\ell_+^2=\sqrt{k^4+\lambdaBB^2}-\lambdaBB$.
The solution may now be written as

\BEQ \label{uib=vib=}
(u_i,v_i)&=&\frac{1}{\sqrt{\tilde\mu L_i}}\,
\frac{k^2\pm(\sqrt{k^4+V^2}+V)}{2k(k^4+V^2)^{1/4}}
\nn\\&\times&
\frac{\sin S(z)}{(\sqrt{k^4+V^2}-V-\ell^2)^{1/4}}.
\EEQ

\ni which also depends on $z$ through $V(z)$. Since $\sin^2S(z)$ oscillates fast,
it can be replaced by $\half$, so
the normalization (\ref{inprodnew})  is achieved by 

\BEQ L_i=\int_{z_i}^0\frac{\d z\,\,(\,\sqrt{k^4+V^2}+V)}
{[(k^4+V^2)(\sqrt{k^4+V^2}-V-\ell^2)]^{1/2}}.
\EEQ

\ni Recalling that $S(z_i)=0$, the hard wall boundary condition $u_i=v_i=s=0$ at 
$z=0$ can be fulfilled provided the phase $S(0)$ is an integer $n$ times $\pi$. 
At given $\ell$ this defines the eigenvalue $k_n$.

\subsubsection{Contribution to the fraction of excited states}

It is instructive to investigate whether these states
cause divergent effects for states localized close to the horizon, 
those with $|z_i|\ll1$. At fixed $k$ and $\ell$, we have

\BEQ 
S'{}^2=
\bk^2-\lambdaBB+\lambdaBB\bA z-\ell^2=\lambdaBB\bA (z-z_i)
\EEQ

\ni with
\BEQ
\bk=(k^4+\lambdaBB^2)^{1/4},\quad \bA =\frac{A(\bk^2-\lambdaBB)}{\sqrt{2}\bk ^2},\quad
\EEQ

\ni and 
\BEQ
\label{ydef} z_i=\frac{\lambdaBB+\ell^2-\bk ^2}{\lambdaBB\bA }.\EEQ

\ni The maximal $\ell$ at a given $z$ is

\BEQ \ell_+^2(z)=\bk^2-\lambdaBB+\lambdaBB\bA z.\EEQ

\ni This brings

\BEQ L_i=
\frac{(\bk^2+\lambdaBB)\sqrt{-z_i}}{\bk^2\sqrt{\lambdaBB\bA}}
=\frac{(\bk^2+\lambdaBB)\sqrt{\bk^2-\lambdaBB-\ell^2}}{\bk^2\lambdaBB\bA}.
\EEQ


From $S(z_i)=0$ we get

\BEQ S(z)=\frac{2}{3}\sqrt{\lambdaBB\bA } (z-z_i)^{3/2}
\EEQ

\ni implying

\BEQ S(0)=\frac{2}{3\lambdaBB\bA}(\bk^2-\lambdaBB-\ell^2)^{3/2}.
\EEQ

So we may set 

\BEQ \frac{\d n}{\d k}=\frac{1}{\pi}\,\frac{\d S(0)}{\d k}\approx 
\frac{2k^3}{\pi\lambdaBB\bA \bk^2 }
\sqrt{\bk ^2-\lambdaBB-\ell^2}.
\EEQ

 We can now calculate at given small $z$

\BEQ &\sum_i& v_i^2(z)
=\int \d k\int_{\ell_c^2}^{\ell_+^2(z)}\d \ell^2\frac{\d n}{\d k}\,v_i^2(z)
\\&=&
\int\d k^2\int_{\ell_c^2}^{\ell_+^2(z)}\d\ell^2 \frac{(\lambdaBB+\bk ^2-k^2)^2}
{16\pi\tilde\mu \bk^2(\lambdaBB+\bk^2)\sqrt{\ell_+^2(z)-\ell^2}} \nn
\EEQ

\ni where we replaced $\sin^2S(z)$ by $\half$. Since the singularity at $\ell_+(z)$ can
be integrated, these states brings no specially large contribution near the horizon $y=0$.
The outcome is of order $\lambdaBB^{3/2}/\xi^{1/2}$, as for the modes near the origin,
so both type of modes bring comparable excitations, $B_e\sim\varepsilon$, as one would expect.

\subsection{States close to the horizon}

States localized close to the horizon can be studied analytically.  
we define 
\BEQ k^4=\ell^4+2\ell^2\lambdaBB(1-\frac{z_i}{\sqrt{2}}),\quad
z_i=\frac{\ell^4-k^4}{\sqrt{2}\ell^2\lambdaBB}
\EEQ

\BEQ \bar z=Cz\EEQ

\ni and supposing $s(z)=f(\bz_i-\bz)$
\BEQ C^4f^{iv}-2(\ell^2+\lambdaBB)C^2f''+2\ell^2\lambdaBB\frac{\bz_i-\bz}{C\sqrt{2}}f=0
\EEQ

\ni so to leading order

\BEQ C=\frac{\lambdaBB^{1/3}}{2^{1/6}}
\left(\frac{\ell^2}{\ell^2+\lambdaBB}\right)^{1/3}\approx
\frac{\lambdaBB^{1/3}}{2^{1/6}}
\left(1-\frac{\lambdaBB}{\sqrt{k^4+\lambdaBB^2}}\right)^{1/3}
\EEQ

\ni The solution is 

\BEQ f_i(z)=\Ai(\bz_i-\bz) \EEQ

\ni which oscillates for $\bz_i<\bz<0$ and decays for $\bz<\bz_i$.
The combination

\BEQ f_i^2(z)=\frac{C\,\Ai^2(\bz_i-\bz)}{2\sqrt{\xi}\int^{\infty}_{\bz_i}\d y\,\,\Ai^2(y)}
\approx  \frac{\pi C\Ai^2(\bz_i-\bz)}{2\sqrt{-\bz_i\xi}}\EEQ

\ni is normalized to $\int\d z \,f_i^2=1$, so this results in

\BEQ (u_i^2,v_i^2)=\frac{(w+V\pm k^2)^2}{4k^2(w+V)}\,f_i^2
\EEQ

The zeros of $\Ai(\bz_i)$ occur at 

\BEQ |\bz_i^{(n)}|=\left(\frac{3\pi n}{2}\right)^{2/3},\quad n
=\frac{2}{3\pi}|\bz_i^{(n)}|^{3/2},
\EEQ

\ni so this yields

\BEQ \d n=\frac{1}{\pi}\sqrt{-\bz_i}\d \bz_i\EEQ

Putting things together yields

\BEQ&& \bar B_e(z)=\frac{1}{2(2\lambdaBB A^2)^{1/6}}\int^0\d \bz_i\,\int_0^\infty\d x
 \left(\frac{X-1}{X}\right)^{1/3}\nn\\&& \frac{(X+1-x)^2}{2X(X+1)}
\Ai^2\left[\bz_i-\left(\frac{X-1}{X}\right)^{1/3}Z\right].
\EEQ

\ni where

\BEQ X=\sqrt{x^2+1},\quad Z=\frac{\lambdaBB^{1/3}z}{2^{1/6}A^{1/3}}.\EEQ

\ni In the $\bar z_i$ integral only the small values are reliable and the result is
not of order unity but of order $\lambdaBB^{-1/6}=4.7\cdot10^{-8}$. Still, 
the result vanishes exactly at $z=0$ before we pass from a sum to the integral.
The derivative is well defined, however,

\BEQ&& \bar B_e'(z)=-\frac{\lambdaBB^{1/6}}{2^{4/3}A^{2/3}}
\int_0^\infty\d x
 \left(\frac{X-1}{X}\right)^{2/3}
\\&& \frac{(X+1-x)^2}{2X(X+1)}
\Ai^2\left[-\left(\frac{X-1}{X}\right)^{1/3}Z\right].
\nn \EEQ

\ni At $z=0$ it takes the value

\BEQ \bar B_e'(0)=-\frac{0.157682\,\lambdaBB^{1/6}}{A^{2/3}},\EEQ

\ni while for $z\to-\infty$ it decays as

\BEQ \bar B_e'(z)=-\frac{9\,3^{1/3}\Gamma(\frac{7}{6})}
{8\,2^{1/4}\sqrt{\pi}}\,\frac{A^{1/2}}{\lambdaBB|z|^{7/6}}
=-\frac{0.03284\,A^{1/2}}{\lambdaBB|z|^{7/6}},
\EEQ

\ni where we used

\BEQ \int_0^\infty\d y\,y^{5/2}\Ai^2(y)=\frac{3^{4/3}\Gamma(\frac{7}{6})}
{4\sqrt{\pi}}= 0.471436.\EEQ

\subsection{Fluctuation energy of the matter field}

Let us calculate the energy density and pressures of the quantum field.
We need the following contributions

\BEQ \label{drhopgen=}
\rho^e_1&=&\frac{\langle\p_t\del\hat\psi^\dagger\p_t\del\hat\psi\rangle}{c^2U}
=\frac{\hbar c}{16\bM^4x^4u}  \langle\p_s\hat\phi^\dagger\p_s\hat\phi\rangle,
\nn\\
\rho^e_2&=&\frac{\langle\p_r\del\hat\psi^\dagger\p_r\del\hat\psi\rangle}{V} =
\frac{\hbar c
\langle(\p_z\hat\phi^\dagger-\frac{y}{\sqrt{2}}\hat\phi^\dagger)
(\p_z\hat\phi-\frac{y}{\sqrt{2}}\hat\phi)\rangle}{16\bM^4x^4v},
\nn\\
\rho^e_3&=&\frac{\langle\del\hat\psi^\dagger L^2\del\hat\psi\rangle}{W^2}=
\frac{\hbar c}{16\bM^4x^4}\langle\hat\phi^\dagger L^2\hat\phi\rangle,\\
\rho^e_4&=&\bm^2\langle\del\hat\psi^\dagger\del\hat\psi\rangle
=\frac{\hbar c\bm^2}{4\bM^2x^2}\langle\hat\phi^\dagger \hat\phi\rangle
 =\frac{3\hbar c\xi}{8\bM^4x^2}\langle\hat\phi^\dagger \hat\phi\rangle,   \nn \\
\rho^e_5&=& \frac{\lambda}{4\hbar c} N\psi_0^2\langle4\del\hat\psi^\dagger\del\hat\psi
+e^{-2i\tilde\mu s}\del\hat\psi^\dagger{}^2+e^{2i\tilde\mu s}\del\hat\psi^2\rangle\nn\\
&=&\frac{3\lambdaB B\hbar c\xi}{16\bM^4x^2}\langle4\hat\phi^\dagger\hat\phi
+e^{-2i\tilde\mu s}\hat\phi^\dagger{}^2+e^{2i\tilde\mu s}\hat\phi^2\rangle.\nn
\EEQ

\ni They determine

\BEQ 
 \rho_e&=&\rho^e_1+\rho^e_2+\rho^e_3+\rho^e_4+\rho^e_5,\nn\\
 p_r^e&=&\rho^e_1+\rho^e_2-\rho^e_3-\rho^e_4-\rho^e_5,\\
p_\perp^e&=&\rho^e_1-\rho^e_2-\rho^e_4-\rho^e_5,\nn
\EEQ

\ni which implies that

\BEQ  T_e=-2\rho^e_1+2\rho^e_2+2\rho^e_3+4\rho^e_4+4\rho^e_5.
\EEQ

\ni The leading terms at $T=0$ are indeed also of order $\varepsilon$,

\BEQ 
\bar\rho^e_1
=\frac{1}{8x^2u}B_e, \quad 
\bar\rho^e_4
=\frac{3}{4}B_e,
\EEQ

\ni the other terms are  smaller by a factor $\lambdaBB$
at least. So

\BEQ \label{repeTe}
\bar\rho_e&=&(\frac{1}{8ux^2}+\frac{3}{4})B_e,\quad p_r^e=p_\perp^e=
(\frac{1}{8ux^2}-\frac{3}{4})B_e,\nn\\
\bar T_e&=&(-\frac{1}{4ux^2}+3)B_e.
\EEQ

\ni This now implies that the sources of the GP equation, (\ref{SBedef}),
and of the Einstein equations, (\ref{rhopBx}) and (\ref{rhoesimpl}), are

\BEQ S_B=0,\quad \bar\rho_e-\frac{3}{4}B_e=\bar p_r^e+\frac{3}{4}B_e=\frac{1}{8x^2}B_e\EEQ

\subsection{Reaction of the metric on the fluctuations}

In order to investigate whether the matching of interior and
exterior solutions can be achieved near the horizon, we investigate
the reaction of the metric to fluctuations of the exact solution
caused by the terms $\rho_e$, $ p_r^e$ and 
$ p_\perp^e$. We set

\BEQ B=1+ B_1,\quad u=1+u_1,\quad v=1+v_1,\quad\xi_1=1+\xi_2,\EEQ

\ni where $B_1$, $u_1$, $v_1$ and $\xi_2$ are of order $\varepsilon$.

\subsubsection{The solvable case A=0}

The $B$-equation becomes to linear order in $\varepsilon$ 

\BEQ
\frac{x^2}{4}B_1''- y B_1'+B_1 =-\frac{2}{3} \bar T_e.
\EEQ

\ni In this section, derivatives are with respect to $y$. The homogeneous solutions are

\BEQ B^{(1)}_1=y,\qquad
B^{(2)}_1=\frac{1}{2x^2}-\frac{3}{2}+\frac{3y}{4}\ln\frac{1-y}{1+y},\EEQ\qquad

\ni and they have a Wronskian

\BEQ 
{\rm W}=B_1^{(1)}B_1^{(2)}{}'- B_1^{(1)}{}'B_1^{(2)}=\frac{1}{x^4}.
\EEQ

\ni The solution for $B_1$ therefore reads

\BEQ 
B_1&=&
b_1y+B_1^{(1)}(y)\int_1^y\d y\frac{8\delta \bar T_m B_1^{(2)}}{3x^2{\rm W}}  \nn\\
&-&B_1^{(2)}(y)\int_1^y\d y\frac{8\delta \bar T_m B_1^{(1)}}{3x^2{\rm W}}=b_1y \\
&+& \frac{4}{3}B_1^{(2)}(y)\int_0^x\d x\,x^3\delta\bar T_m 
-  \frac{4y}{3}\int_0^x\d x\frac{x^3}{y}    \delta\bar T_m  B_1^{(2)}, \nn
\EEQ

\ni with $b_1$ an integration constant.
The last expression exhibts regularity at the origin $x=0$, even when $\delta T_m$ 
has a $1/x$ singularity, as we discussed above. 

$v_1$ may be solved from the $^1_1$- Einstein equation, 


\BEQ v_1&=& \frac{x^2}{y^2}\left[\frac{u_1-x^2yu_1'}{3x^2}
- \frac{4}{3}\delta p_r^m +\half(B_1-yB_1')\right],\EEQ

\ni after which $u_1$ satisfies

\BEQ 
x^2  u_1''-4y u_1'+\frac{4}{x^2}u_1=s\EEQ


\ni with source term 

\BEQ s=-4\bar\rho_1+14\bar\rho_4-\frac{x^2}{2y}(\bar\rho_1'-\bar\rho_4')
\EEQ

\ni The homogeneous solutions $P/x$ and $Q/x$
involve the associated Legendre functions $P\equiv P_1^{i\sqrt{3}}$, 
$Q\equiv Q_1^{i\sqrt{3}}$, 

\BEQ P(y)=\left(\frac{1+y}{1-y}\right)^{\frac{i}{2}\sqrt{3}}(1+\frac{iy}{\sqrt{3}}),\quad
Q(y)=P^\ast(y)
\EEQ

\ni The solution then reads 

\BEQ \label{u1=}
u_1&=&\frac{Q}{x}\int_1^y\d y \frac{Ps}{x{\rm W}}-\frac{P}{x}\int_1^y\d y\frac{ Qs}{x{\rm W}}
\nn\\
&=&\frac{P}{x}\int_0^x\d x\frac{ Qs}{y{\rm W}}-
\frac{Q}{x}\int_0^x\d x \frac{Ps}{y{\rm W}}\EEQ

\ni with the Wronskian

\BEQ {\rm W}=PQ'-QP'=-\frac{8i}{\sqrt{3}\,x^2}.
\EEQ

\ni It is imaginary, so $u_1$ is real. (\ref{u1=}) is regular at $x=0$, so
no homogeneous solutions can be added. 
It now follows that $v_1(y)$ diverges as $c_{-2}/y^2$ near the horizon $y=0$.
We get

\BEQ c_{-2}=\frac{u_1(0)}{3}-\frac{4}{3}[\bar\rho_1(0)-\bar\rho_4(0)]+\half B_1(0)
\EEQ

\ni with 

\BEQ u_1(0)=\frac{i\sqrt{3}}{8}\int_0^1\d y\,(Q-P)xs
\EEQ

\ni For $A=0$ one has

\BEQ \bar\rho_1=\frac{\rho_a}{x}, \quad 
\bar\rho_4=6\rho_ax
\EEQ

\ni with $\rho_a=2.80959\cdot10^{-15}M_9^3$ a positive amplitude. 
This gives $u_1(0)=13.6518\, \rho_a$, while $B_1(0)=-496\rho_a/45$. Together they
yield

\BEQ c_{-2}=5.70616 \rho_a,\EEQ

\ni which is positive, and showing an upturn of $v$ in the narrow
region $y\sim\sqrt{\varepsilon}$ near the horizon.

\subsubsection{The general case $A>0$}

When $A>0$, the $B_1$ equation gets coupled to the $u_1,v_1$ equations.
No explicit solution of the linearized problem has been found. 
Inspection of the equations near $y=0$ reveals that now a singularity

\BEQ
v_1=\frac{c_{-2}}{y^2}
+O(y^0),\quad
 B_1=-\frac{Ac_{-2}}{2y}+O(y^0),\quad 
\EEQ

\ni is allowed. From the above case $A=0$ it is to be expected that $c_{-2}$
remains positive, so at $A>0$ also $B_1$ is singular. The signs are the
ones expected for approaching the vacuum: $B$ decays while $v$ increases,
towards its high peak slightly beyond the horizon. 
In retrospect, the induced decay of $B$ also indicates that $A>0$ is the typical case, 
rather than the no-hair value $A=0$.
This important fact gives hope that a self-consistent treatment near the
horizon achieves to match the exact solution in the interior with
the deformed Schwarzschild metric in the exterior.
It remains an open problem to consider this behavior in a self-consistent way.

\section{Conclusion}
We have questioned the general wisdom that static BHs have all their mass in the 
center and that its interior cannot be described by present theories based on 
General Theory of Relativity. Estimates show that a picture of closely packed H
atoms naturally applies to the supermassive BH's in the center of galaxies,
$M\sim 10^9M_\odot$. We therefore attempt to describe them as 
more or less normal objects like stars.

We present within the Relativistic Theory of Gravitation (RTG),
an exact solution for a BH, of which the interior is governed
by quantum matter in its Bose-Einstein condensed phase.
Powerlaw singularities occur at the origin, that get absorbed in the Riemann description
of the metric. Elsewhere, the solution is regular. The redshift at the horizon is
finite, though of the order $1/\mub \sim 10^{14}$.

This solution is still to be matched with the Schwarzschild metric, which
near the horizon is deformed in RTG.  We have carefully derived the 
complete fluctuation spectrum about the groundstate.
The matching of the inner metric with the outer metric at the horizon
has been considered in a first order perturbative approach, which shows
an enhancement effect near the horizon. The full problem still has to be
carried out,  and it has to be done self-consistently. 
It remains as a task for future to show that this indeed leads to a proper
decay of matter and behavior of the metric near the horizon.

Our BH is a quantum fluid confined by its own gravitation.
It puts forward that a BH is just an intense gas cloud, without an event horizon, as 
was also deduced from the Schild et al. observations ~\cite{Schildetal06,Schildetal09}.
In the interior, time keeps its standard role.
No Planckian physics is involved; Hawking radiation is absent
and Bekenstein-Hawking entropy plays no role. In our zero-temperature situation 
the entropy of the quantum field vanishes. Because the Schwarzschild 
singularity is cut off by the bimetric coupling, there is no connection with
any form of quantum gravity,  even though the redshift at the horizon is of
order $10^{14}$.

Schild et al. ~\cite{Schildetal06,Schildetal09}
have explained their observations in term of a magnetic dipole moment
of the black hole, a ``hair''.
Our BH also has one ``hair'', the binding energy, expressed
as $E_{\rm bind}=Nmc^2-Mc^2=(1-\nu)Nmc^2$.  Here $\nu$ can take any value
below $1/2\sqrt{3}=29\%$.
We confirmed that the previously derived solution of the Gross-Pitaevskii equation 
with a free parameter, $A$ or $\nu(A)$, 
shows up also as a zero mode in the fluctuation spectrum.
As one would expect for a classical theory of gravitation, when the quantum matter in the
BH has reached a certain groundstate, the classical metric allows the system
still to go to a lower energy state.
Indeed, the passage of celestial bodie will induce oscillations in the metric
and emission of gravitational waves, which, upon re-equilibration, increase the binding energy,
finally up to $100\%$ of the rest energy of its constituents, $Nmc^2$.
In that final state the mass is completely balanced by the binding energy, 
making it look like a zero mass object.
In its stable state, the BH has a fraction strictly-less-than-one-half of the ground state energy 
of the constituents, so the major part of the zero-point energy has to be emitted in radiation.
This property may explain the enormous luminosity of quasars.

It has been assumed that renormalization couples to the matter field density to the
curvature scalar with a strength $\xi$. This parameter is chosen appropriately in a first 
step of renormalization of the scalar field theory with a large quartic coupling.  
The value $\xi =2m^2M^2/3m_P^4\sim 3\cdot 10^{55}(M/10^9M_\odot)^2$ 
allows an exact and explicit solution of the interior metric, any other value would be
inconsistent. The fluctuation spectrum is well defined, and various very large or very small
numbers finally combine into reasonable prefactors. It is noticed that the leading 
corrections of the matter field are of order $\sqrt{na_s^3}$, as it happens for
in Bose-Einstein condensation in a box. 
The physical reason for this, diluteness of the gas of H-atoms,  
was put forward in the introduction.  For this reason we expect that the field theory
for supermassive BHs with $M\gg10^9M_\odot$ can be renormalized 
perturbatively after the first step that fixes the leading value of $\xi$.

We have set out the lines for studying the fluctuation spectrum near the horizon.
It is left as a task for future to show that they indeed fluently connect the
empty space metric of the exterior (i. e. the deformed Schwarzschild metric)
with our exact solution for the metric in the interior.

An important question is whether formation of realistic supermassive BHs
brings the matter indeed in or near the Bose-Einstein condensed groundstate.
This would require the study of the finite temperature situation.
Extension to finite temperatures, not presented here,
will exhibit a $T^{3/2}$ fraction of thermal atoms.
Also the stability of the solution needs to be studied.

Calculation of the normal mode spectrum may lead to predictions that deviate from
the ones of GTR; this spectrum may be observed in the foreseeable future.

We failed to apply our approach to GTR, technically because it lacks
compensation for the $1/U$ terms, that in this solution are truly singular at the origin. 
If no other solution exists for the considered physical situation, 
GTR must be abandoned and replaced by another theory, RTG
being the first candidate. In view of its smaller symmetry group, this may have far
reaching consequences for singularities in classical gravitation -- they would probably be
regularized -- and for quantum approaches to gravitation, 
since the primary space-time, Minkowski space-time, needs no quantization.

\section*{Acknowledgements} 
Th. M. N. has benefited from discussion
with Ugo Moschella and Bernard Nienhuis.

\end{document}